\pdfoutput=1
\documentclass[a4paper,fleqn,usenatbib]{mnras}

\usepackage[T1]{fontenc}
\usepackage{ae,aecompl}
\usepackage{graphicx}
\usepackage{physics}
\usepackage{amsmath}

\usepackage{graphicx}	
\usepackage{amsmath}	
\usepackage{amssymb}	
\newcommand{\msun}{{\rm M}_\odot}
\usepackage{newtxtext,newtxmath}

\usepackage[normalem]{ulem}

\title[Stability analysis of supermassive stars.]{Stability analysis of supermassive primordial stars: a new mass range for general relativistic instability supernovae. }

\author[C. Nagele et al.]{
Chris Nagele,$^{1}$\thanks{E-mail: chrisnagele.astro@gmail.com}
Hideyuki Umeda,$^{1}$
Koh Takahashi, $^{2,3}$
Takashi Yoshida,$^{4}$
Kohsuke Sumiyoshi$^{5}$
\\
$^{1}$Department of Astronomy, Graduate School of Science, the University of Tokyo, Tokyo, 113-0033, Japan\\
$^{2}$Astronomical Institute, Graduate School of Science, Tohoku University, Sendai, 980-8578, Japan\\
$^{3}$Max Plank Institute for Gravitational Physics (Albert Einstein Institute), D-14476 Potsdam, Germany\\
$^{4}$Yukawa Institute for Theoretical Physics, Kyoto University, Kyoto 606-8502 Japan\\
$^{5}$National Institute of Technology, Numazu College, Ooka 3600, Numazu, Shizuoka 410-8501, Japan
}

\date{Accepted XXX. Received YYY; in original form ZZZ}

\pubyear{2022}

\begin{document}
\label{firstpage}
\pagerange{\pageref{firstpage}--\pageref{lastpage}}
\maketitle

\begin{abstract}
    Observed supermassive black holes in the early universe have several proposed formation channels, in part because most of these channels are difficult to probe. One of the more promising channels, the direct collapse of a supermassive star, has several possible probes including the explosion of a helium-core supermassive star triggered by a general relativistic instability. We develop a straightforward method for evaluating the general relativistic radial instability without simplifying assumptions and apply it to population III supermassive stars taken from a post Newtonian stellar evolution code. This method is more accurate than previous determinations and it finds that the instability occurs earlier in the evolutionary life of the star. Using the results of the stability analysis, we perform 1D general relativistic hydrodynamical simulations and we find two general relativistic instability supernovae fueled by alpha capture reactions as well as several lower mass pulsations, analogous to the puslational pair instability process. The mass range for the events (2.6-3.0 $\times 10^4$ $\msun$) is lower than had been suggested by previous works (5.5 $\times 10^4$ $\msun$) because the instability occurs earlier in the star's evolution. The explosion may be visible to, among others, JWST, while the discovery of the pulsations opens up additional possibilities for observation.

\end{abstract}



\suppressfloats

\section{Introduction}
\label{introduction}

For much of the history of astronomy, the post recombination early universe has been inaccessible to observation. While this state of affairs still holds broadly, inroads are being made. Observers have detected a star at redshift 6 \citep{welch2022}, galaxies as early as redshift 11 \citep{oesch2016}, long gamma ray bursts (GRB) at redshifts 8 and 9 \citep{tanvir2009,cucchiara2011}, and quasars at redshifts 6 and 7 \citep{mortlock2011,wu2015,banados2018,matsuoka2019,wang2021}. These high redshift observations have greatly increased our knowledge of the early universe, but they also raise questions--- most notably, where did the high redshift quasars and their supermassive black hole (SMBH) engines come from?

Several theories have been put forward to explain the existence of SMBHs so soon after the big bang \citep[e.g.][]{rees1984,inayoshi2020}. These include, but are not limited to, the direct collapse scenario \citep{bromm2003,lodato2006}, super-Eddington accretion onto solar mass black holes \citep{haiman2001,madau2014,volonteri2015}, runaway stellar mergers in nuclear star clusters \citep{devecchi2009,katz2015,das2021}, and rapid mergers of either primordial \citep{bean2002} or astrophysical black holes \citep{omukai2008}. Of these scenarios, the direct collapse scenario may be the easiest to test observationally; Population III (Pop III) stars are the first generation of stars and because of the lack of metals in the primordial gas out of which they form, a small fraction of these stars may be supermassive stars (SMS). These SMSs are potentially directly observable by JWST and, with the assistance of strong lensing, by Euclid \citep{surace2018,surace2019,vikaeus2022}. Some of these Pop III SMSs may undergo a general relativistic instability supernova (GRSN)\citep{chen2014,nagele2020}, and \citet{whalen2013} and \citet{moriya2021} have showed that the GRSN should be visible to JWST \citep{gardner2006,kalirai2018}, even at high redshift. The GRSN may leave other observable imprints on the proto-galaxy, specifically Pop III starbursts, metal enrichment or X-ray emission from the remnant \citep{Whalen2013ApJ...774...64W,Johnson2013ApJ...775..107J,Whalen2013ApJ...777...99W} . Pop III SMSs may also produce GRBs \citep{sun2017} and gravitational waves \citep{shibata2016,li2018} visible to future detectors such as THESEUS \citep{amati2018}, LISA \citep{amaro-Seoane2017}, and DECIGO \citep{kawamura2011}.

Primordial star formation in the early universe is commonly thought to be disrupted in two scenarios. First, if a strong UV-Lyman Werner background exists,  H$_2$ molecules will be photodissociated and the the gas will have no way to cool below $10^4$ K given the primordial composition and thus cannot fragment to form Pop III stars \citep{Dijkstra2008MNRAS.391.1961D,Agarwal2012MNRAS.425.2854A,Latif2014MNRAS.443.1979L}. Next, supersonic baryon-dark matter streaming can prevent fragmentation and thus Pop III star formation \citep{Latif2014MNRAS.440.2969L,Schauer2017MNRAS.471.4878S,Hirano2017Sci...357.1375H}. In both of these scenarios, the gas does not collapse until it reaches a mass of around $10^7$ $\msun$, after which gravitational instability leads to the formation of a single object with mass up to $10^5$ $\msun$ \citep{Wise2008ApJ...685...40W,latif2013,regan2017}. 

Two subsequent scenarios are studied, the first assumes that accretion onto the central protostar eventually terminates (e.g. from UV feedback during a quiescent phase \citealt{Sakurai2015MNRAS.452..755S}), after which the star enters the main sequence, and it is thus referred to as the non accreting scenario \citep{fuller1986,montero2012,chen2014,nagele2020,woods2020}. The second, where the protostellar accretion occurs at a constant rate, is known as the accreting scenario \citep{hosokawa2012,hosokawa2013,schleicher2013,umeda2016,woods2017,hammerle2018a}. One subtlety is that if the early accretion occurs too rapidly, hydrogen burning cannot ignite and the star will become rotationally supported, eventually collapsing to a black hole \citep{shibata2002}. Even if the star does not become rotationally supported, it is thought to rotate near the $\Omega-\Gamma$ limit \citep{hammerle2018}.

Considering the non accreting scenario, \citet{chen2014} discovered that for a small mass range around 55000 $\msun$ the general relativstic instability \citep{chandrasekhar1964} occurs when the SMS has a large reserve of helium, and the subsequent collapse triggers explosive helium burning which disrupts the star in a GRSN. Previously, we investigated this phenomenon using a post Newtonian stellar evolution code (as in \citealt{chen2014}) followed by a general relativistic hydrodynamical code adapted from pair instability and core collapse supernovae simulations \citep{nagele2020}. We found slightly different results from \citet{chen2014}, but did confirm that a GRSN was possible. In \citet{nagele2020}, our main difficulty had been determining when to connect the two codes. In this paper, we perform a general relativistic stability analysis for a better determination, and using this we find a significantly altered mass range for the GRSN.

This paper is organized as follows. In Sec. \ref{methods} we discuss the various codes and methods of analysis as well as the numerical models. In Sec. \ref{results_comp} we compare the results of the stability analysis to other methods. In Sec. \ref{results_explo} we present simulations of GRSNe using the results of the stability analysis and compare the resulting ejecta to observations of metal poor stars. In Sec. \ref{results_col}, we discuss how this change might effect the neutrino emission of collapsing SMSs. Finally, we conclude with a discussion in Sec. \ref{discussion}.

\begin{figure*}
    \centering
    \includegraphics[width=2\columnwidth]{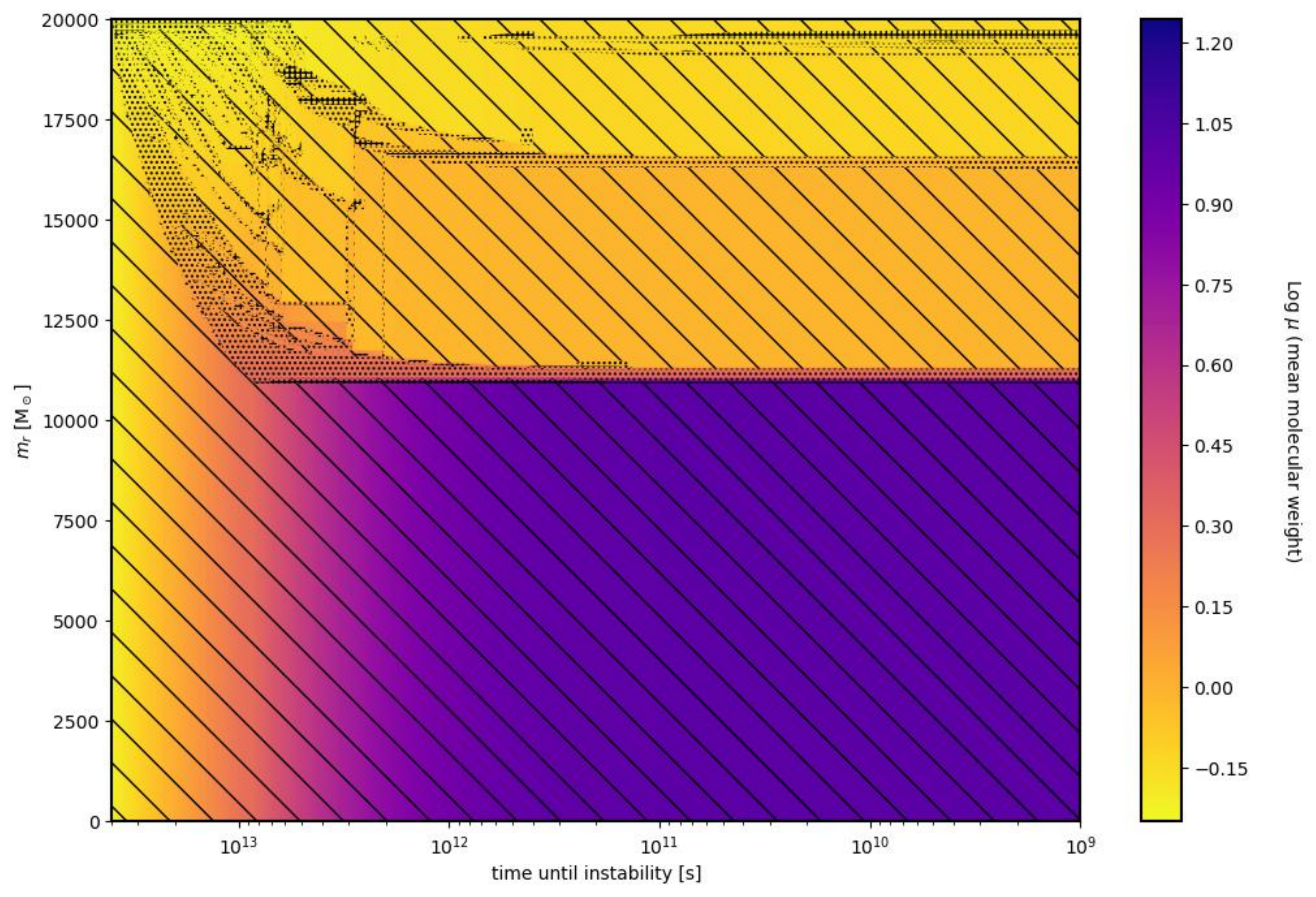}

    \caption{Kippenhahn diagram for the lowest mass model ($2 \times 10^4$ $\msun$). The color shows the log of the mean molecular weight. Diagonal hatches show convective regions, dotted hatches show non convective regions, and cross hatches show semi-convective regions. Note that the time-step of HOSHI at this point in the evolution is roughly $10^9$ s. }
    \label{fig:kipp}
\end{figure*}

\begin{figure}
    \centering
    \includegraphics[width=\columnwidth]{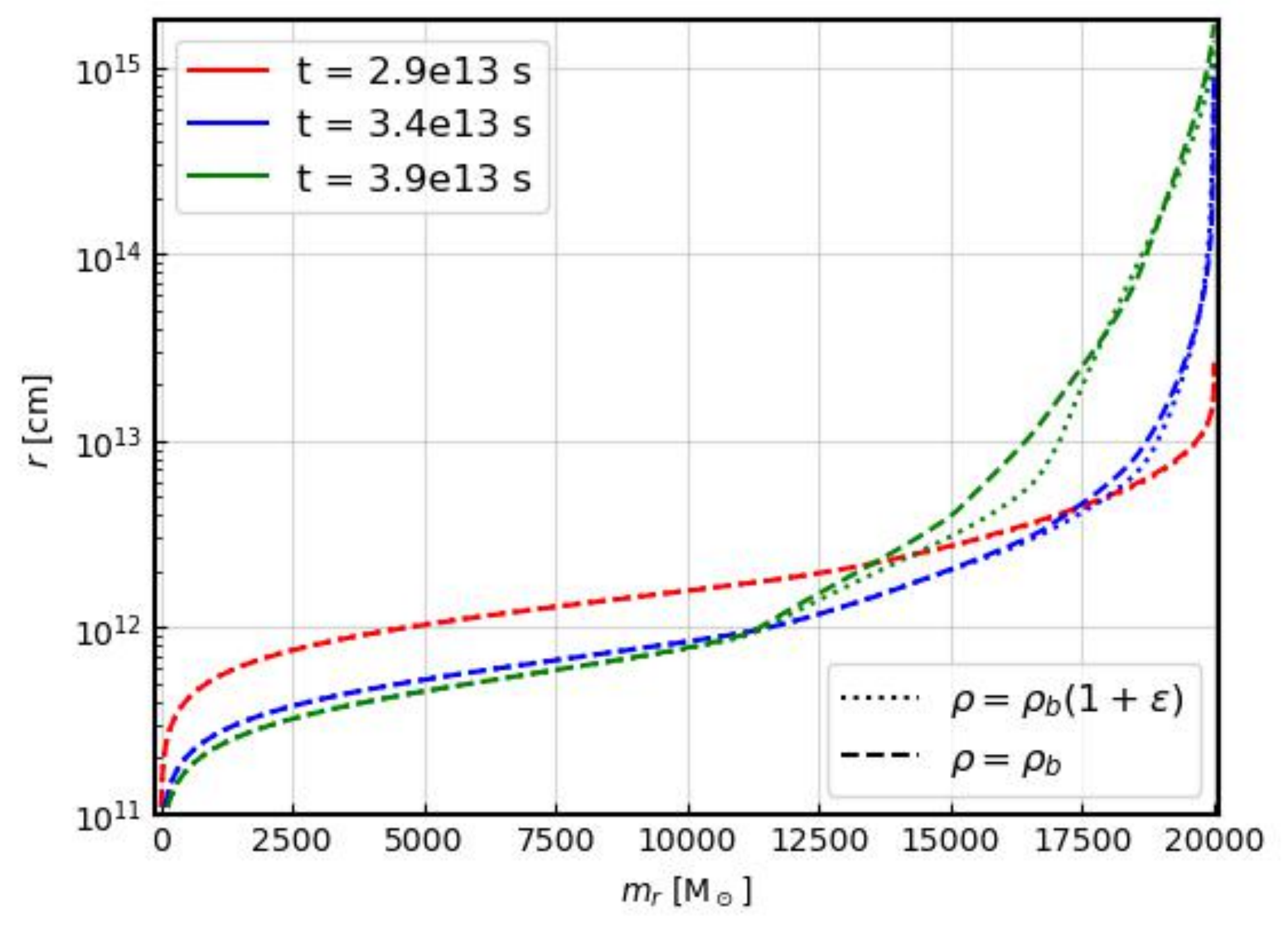}

    \caption{Comparison of radial time snapshots for the 2 $\times 10^4$ $\msun$ model in the HOSHI code used in this work (dotted lines) and the one used in \citet{nagele2020} (dashed lines). The inclusion of internal energy to the GR density when evaluating the PN TOV equation causes a more compact inner envelope, while the outer envelope remains largely unchanged.}
    \label{fig:HOSHI_r}
\end{figure}

\section{Methods}
\label{methods}

The GRSN occurs when a non accreting SMS experiences the general relativstic (GR) instability during helium burning. The star then contracts before rapidly burning a fraction of its helium and then exploding. 

To model this phenomenon, we first evolve the star from just before the onset of nuclear burning, using a stellar evolution code (Sec. \ref{methods_HOSHI}). At each timestep in the evolutionary calculation, we use a general relativistic stability analysis to check if the star is stable (Sec. \ref{methods_stab}). When the star becomes unstable according to the stability analysis, we map the evolutionary model to a hydrodynamical code and follow the evolution until the star collapses, explodes, or pulsates (Sec. \ref{methods_HYDnuc}). 

\subsection{Stellar Evolution}
\label{methods_HOSHI}

The HOSHI code is a stellar evolution code which, in this work, solves the hydrodynamical equations in post Newtonian (PN) gravity using a Henyey type implicit method (Newton-Raphson method), taking the density, entropy, radius, and luminosity as the independent variables \citep{takahashi2016,takahashi2018,takahashi2019,yoshida2019,nagele2020}. Although we describe HOSHI as a stellar evolution code (it solves the stellar structure equations, including energy transport), the code is able to follow hydrodynamical evolution to a degree, though it does not include a shock capture scheme. This, along with the lack of full GR, is why we require HYDnuc to model the GRSN. HOSHI includes a nuclear reaction network (52 isotopes), neutrino cooling, mass loss (though it is minimal for non rotating primordial stars), and rotation. The equation of state includes contributions from photons, averaged nuclei, electrons, and positrons. We use the Rosseland mean opacity of the OPAL project \citep{Iglesias1996ApJ...464..943I} and solve the Saha equation to determine the ionization of hydrogen, helium, carbon, nitrogen, and oxygen. Several effects of convection are modeled including chemical mixing through diffusion in convective and semi-convective regions. Finally, energy transfer due to convection is treated according to 1D mixing length theory.  

The star is initiated with Log $T_c$ $<$ 7.7 in a high entropy state relative to ZAMS and has a primordial composition except for deuterium which we have removed to avoid the proto-stellar burning phase, which is not important for the evolution (see e.g. \citealt{Hosokawa2009ApJ...691..823H}). Because we initialize the star as a supermassive protostar, its structure is very nearly that of an $n=3$ polytrope. The star will undergo a moderate period ($\sim 10^{12}$ s) of contraction. The p-p chain is insufficient to stop this contraction and so the star must reach a central temperature around Log $T_c$ = 8.2 at which point the triple alpha reaction produces enough carbon to ignite the CNO cycle, which stabilizes the star. The star continues to be supported by the CNO cycle until hydrogen is exhausted, at around one million years. The star then contracts, with the central temperature increasing to Log $T_c$ = 8.5 before helium burning stabilizes the star. The stars in this paper become unstable to the GR radial instability (Sec. \ref{methods_stab}) in late hydrogen burning or in helium burning. Fig. \ref{fig:kipp} shows a Kippenhahn diagram for the M=$3\times 10^4$ $\msun$ model. At the beginning of hydrogen burning, the star is nearly fully convective (diagonal hatches), though towards the end of this period, a core and envelope form. The core remains convective until the instability while the envelope has several convective layers separated by areas of non convection (dots) or semi-convection (crosses). Hydrogen shell burning proceeds in the non convective layer just above the core. The onset of the instability is sudden, as is the case with the pair instability, and is not precipitated by major changes in the composition (see color-bar) as in the case with core collapse.

HOSHI uses the first order PN approximation to the Tolman Oppenheimer Volkoff (TOV) equation. However, unlike in \citet{chen2014,nagele2020}, we include the correction of energy to density:
\begin{equation}
    \rho = \rho_0 \bigg(1+\frac{\epsilon}{c^2}\bigg)
    \label{eq:epsilon}
\end{equation}
where $\rho_0$ is the baryonic density and $\epsilon$ is the specific internal energy in units of ergs g$^{-1}$. We use the convention that the rest mass energy due to the mass excess of isotopes is included in the internal energy. $\epsilon/c^2$ is between 0.01 and 0.001 throughout the star and its inclusion is necessary to correctly model the SMS envelope. Fig. \ref{fig:HOSHI_r} shows the results of the stellar evolution calculation compared to \citet{nagele2020}.   

Beyond correctly modeling the SMS envelope, the inclusion of internal energy is necessary for consistency between HOSHI and HYDnuc. If we only use the baryonic density when calculating the TOV PN terms in HOSHI, then the $\mathcal{O}(1\%)$ difference in the gravity will perturb the star too strongly in HYDnuc, causing most models--- even some stable configurations--- to collapse to black holes. 

This work contains twenty supermassive stars with different masses, where we have chosen those masses to be centered around the explosion window. The explosion window is heavily dependent on when the GR instability triggers, as an explosion is only possible if the instability occurs during helium burning.

\subsection{GR Stability Analysis}
\label{methods_stab}

\begin{figure}
    \centering
    \includegraphics[width=\columnwidth]{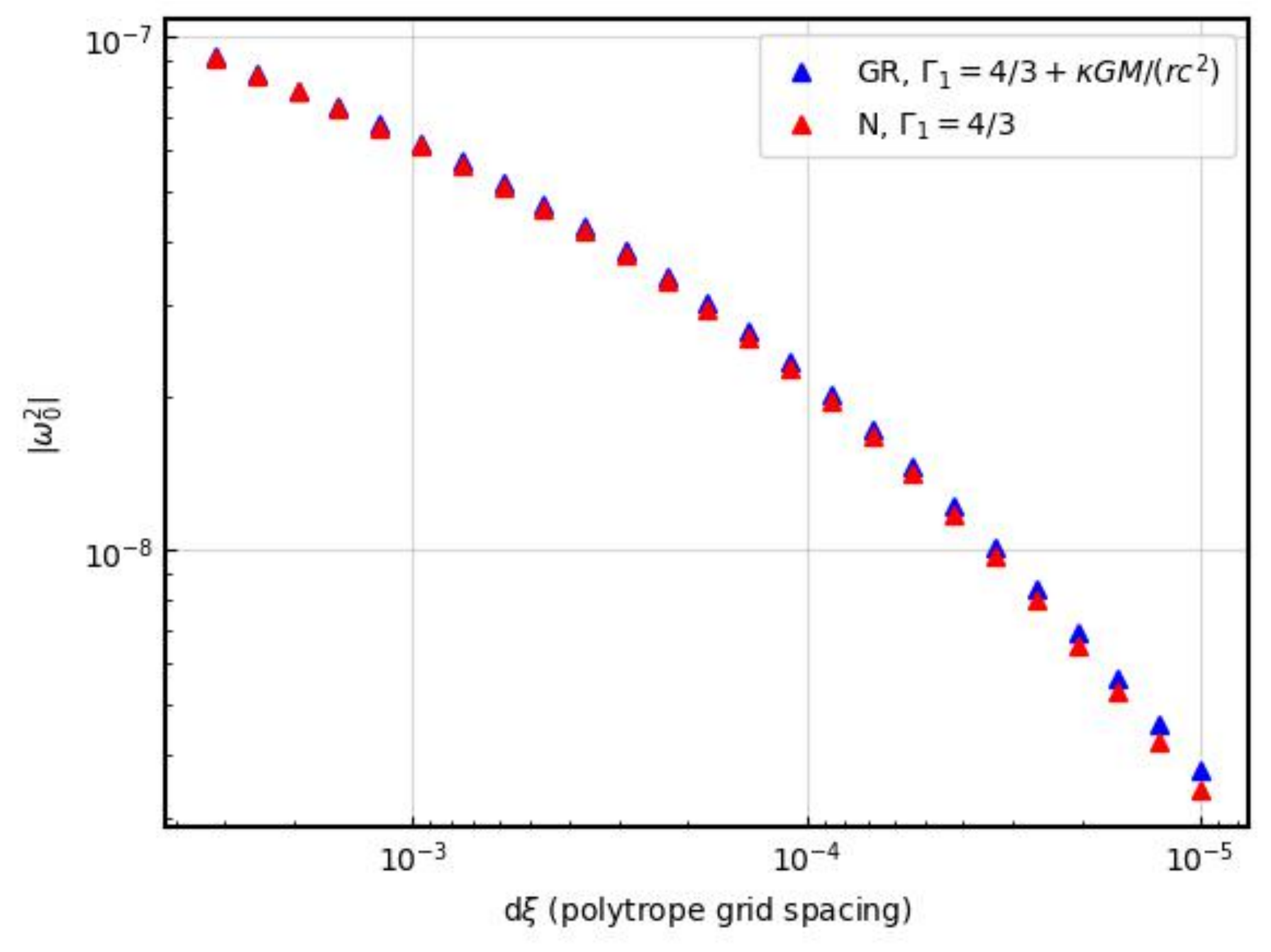}

    \caption{Numerical convergence of the stability analysis code for numerical polytropes with spacing d$\xi$. These polytropes are evaluated at $\Gamma_1=4/3$ for the Newtonian analysis and $\Gamma_1=4/3+\kappa GM/(rc^2) $ for the GR analysis so that we should find $\omega_0^2 = 0$. Thus, the y axis can be regarded as the error associated with our method, and its amplitude decreases for increasing numerical resolution. The number of mesh points in HOSHI is variable, but is usually between 1000 and 2000, translating to somewhere on the left side of this plot. }
    \label{fig:nconv_inst}
\end{figure}

The question of when a SMS collapses due to the GR instability is a challenging one because the collapse takes place on timescales far smaller than the typical timestep of an evolutionary calculation. In addition, the TOV equation lacks the dynamical GR terms which make the collapse so rapid. It would likely require a relativistic stellar evolution code to fully address the problem of SMS collapse. In \citet{nagele2020}, we relied on the PN stellar evolution code to determine when collapse would occur, and our results were in rough agreement with those of \citet{chen2014}. In this work we perform a stability analysis on the normal modes of radial perturbations of a star in GR \citep{chandrasekhar1964}. 

Consider an infinitesimal, radial, Lagrangian perturbation which varies in time ($t$) as $\xi \propto e^{i\omega t}$ for $\omega^2 \in \mathbb{R}$. In Newtonian gravity, this perturbation obeys the equation \citep{shapiro1983}

        \begin{equation}
            \label{eq_N}
            \dv{}{r} \; \bigg[ \Gamma_1 \frac{P}{r^2} \dv{}{r} \; (r^2\xi)\bigg] - \frac{4}{r}\dv{P}{r} \xi +\omega^2\rho_0\xi = 0
        \end{equation}
where $P$ is the pressure, $r$ is the radius, $\rho_0$ is the baryonic density and $\Gamma_1$ is the local adiabatic index at constant entropy (s):
\begin{equation}
    \Gamma_1 = \pdv{\ln P}{\ln \rho_0}  \bigg\rvert_s.
\end{equation}

Solving this differential equation for $\xi$ and $\omega$ is a Sturm–Liouville eigenvalue problem. Finding any solution with $\omega^2 < 0$ is a sufficient condition for instability, as the motion of the perturbation will be exponential. Sturm–Liouville equations have the property that a sequence of solutions exist 
\begin{equation}
    \omega^2_0<\omega^2_1<\omega^2_2<...
    \label{eq_oms}
\end{equation}
corresponding to $\xi_i$s where $i$ is the number of nodes in the perturbation. Because of the above property, a necessary condition for instability is

\begin{equation}
    \omega^2_0 < 0.
\end{equation}
The corresponding equation in GR is \citep{chandrasekhar1964}
        \begin{equation}
            \nonumber
            e^{-2a-b}\dv{}{r} \; \Big[ e^{3a+b}\Gamma_1 \frac{P}{r^2} \dv{}{r} \; (e^{-a}r^2\xi)\Big] - \frac{4}{r}\dv{P}{r} \xi +e^{-2a-2b}\omega^2(P+\rho c^2)\xi 
        \end{equation}     
        \begin{equation}
            \label{eq_GR}
            - \frac{8\pi G}{c^4}e^{2b}P(P+\rho c^2)\xi - \frac{1}{P+\rho c^2} \bigg(\dv{P}{r}\bigg)^2 \xi = 0
        \end{equation}
where $a,b$ are the metric coefficients as in \citet{haemmerle2020} and the density is defined in Eq. $\ref{eq:epsilon}$.

\begin{figure}
    \centering
    \includegraphics[width=\columnwidth]{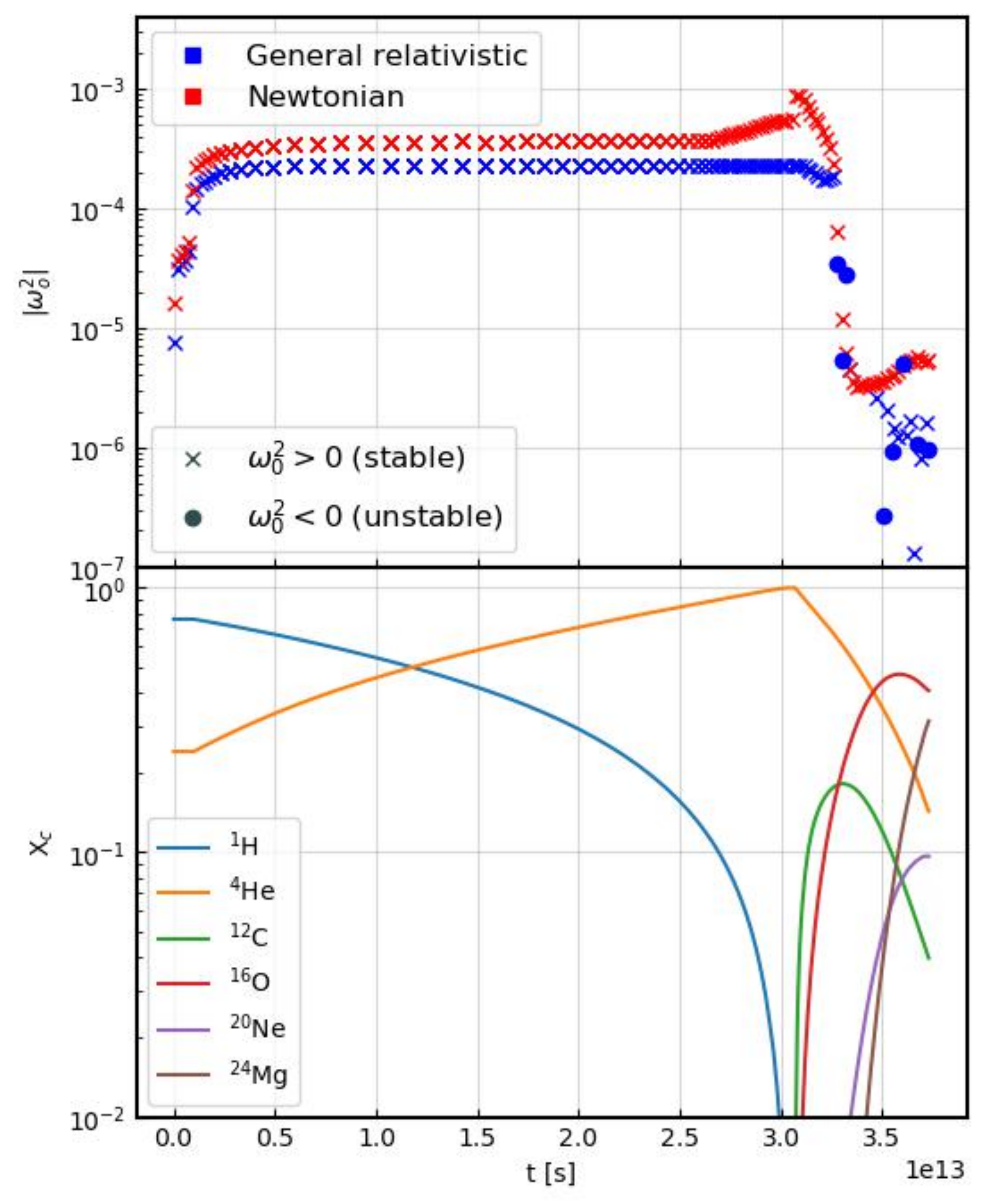}

    \caption{GR stability analysis (Sec. \ref{methods_stab}) applied to the results of the 3 $\times 10^4$ $\msun$ model in HOSHI. Upper panel --- amplitude of the fundamental mode frequency as a function of time for the General Relativistic (blue) and Newtonian (red) analyses. Unstable models, having a negative value of frequency squared, are denoted by filled circles while stable models are denoted by crosses. Lower panel --- Central mass fraction of various isotopes as a function of time. The first instability occurs during helium burning. }
    \label{fig:insta_hoshi}
\end{figure}

\begin{figure}
    \centering
    \includegraphics[width=\columnwidth]{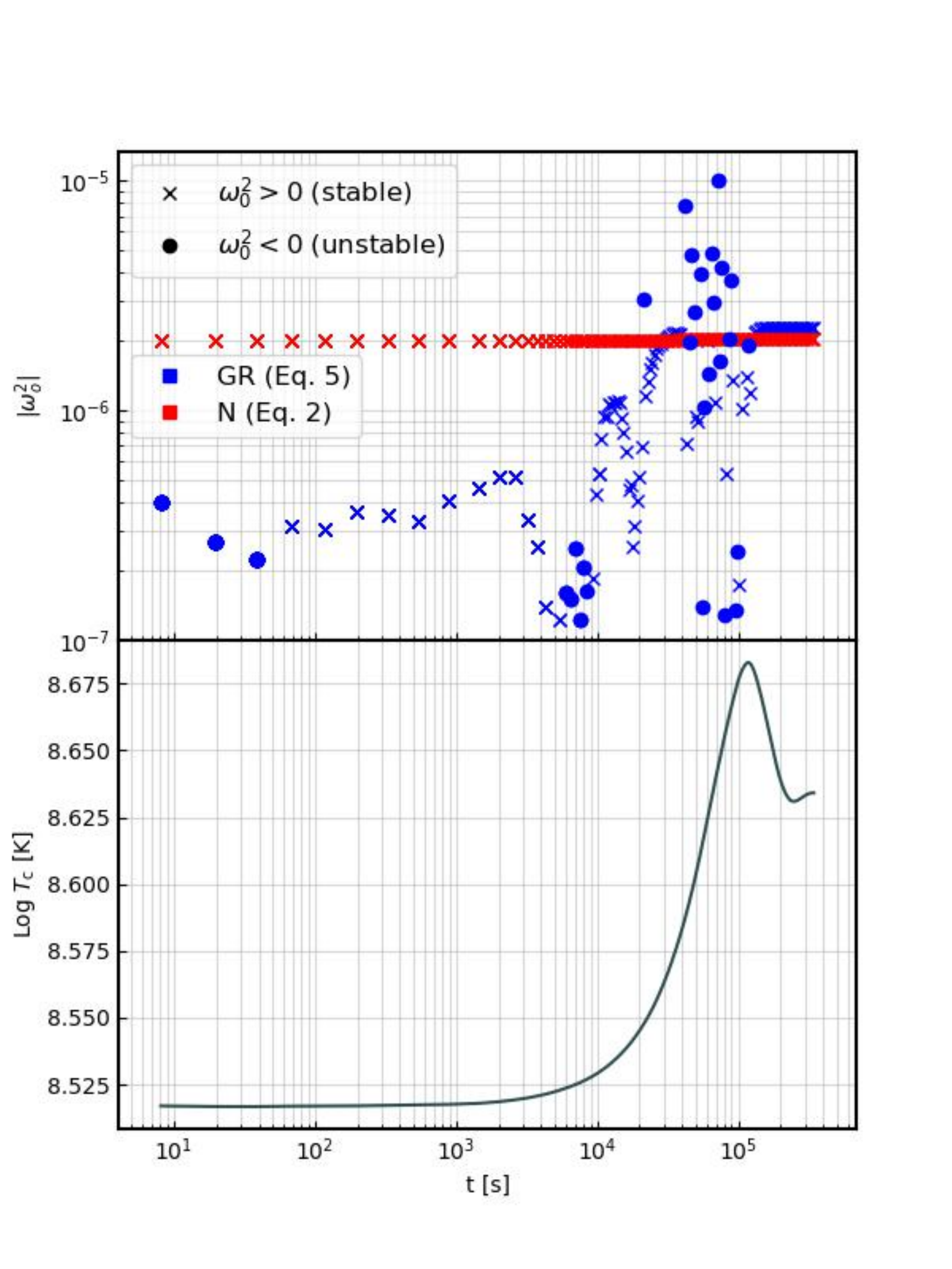}

    \caption{Stability analysis performed on the results of the HYDnuc calculation for the first unstable model of 2 $\times 10^4$ $\msun$. Upper panel --- stability analysis. Lower panel --- central temperature. }
    \label{fig:insta_hydro}
\end{figure}

\begin{figure*}
    \centering
    \includegraphics[width=2\columnwidth]{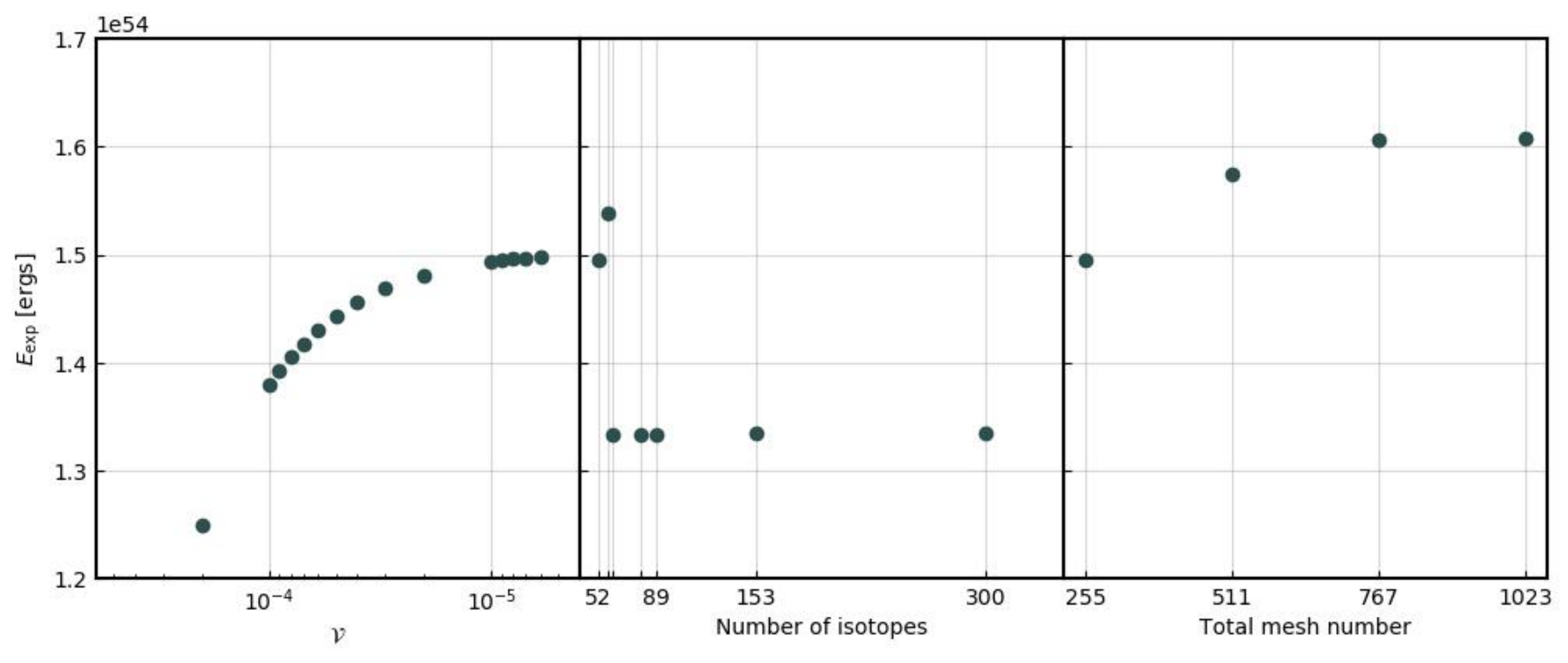}

    \caption{Numerical convergence of explosion energy for three parameters of HYDnuc for the 3 $\times 10^4$ $\msun$ model. Left panel --- dependence on $\mathcal{V}$ (Eq. \ref{eq_varlim}) with total mesh point number = 255 and isotope number = 52. Middle panel ---  dependence on number of isotopes in the nuclear network (52, 58, 61, 79, 89, 153, 300; see Appendix B, Table \ref{tab:networks}) with total mesh point number = 255 and $\mathcal{V} =5\times 10^{-4}$. Right panel ---  dependence of explosion energy on number of mesh points, with $\mathcal{V} = 5\times 10^{-4}$ and isotope number = 52.}
    \label{fig:exp_nconv}
\end{figure*}

In order to solve Eqs. \ref{eq_N},\ref{eq_GR} for $\omega^2_0$, we adopt an iterative method similar to that outlined in exercise 6.11 of \citet{shapiro1983} (for a detailed discussion, see Appendix A). For a given stellar profile ($r, m_r(r), \rho(r), P(r)$), we choose a value of $\omega^2$ and integrate Eq. \ref{eq_GR} (or Eq. \ref{eq_N}, the procedure is identical) to find $\epsilon$. This is done twice, once starting from the center and once starting from the stellar surface. If the two solutions match at some test radius, then $\omega^2$ is a solution to Eq. \ref{eq_GR}. If not, we repeat the procedure, extrapolating new values of $\omega^2$ based on the Wronskian of the two solutions at the test radius. Once a solution is found, we check if it is the fundamental mode by determining the number of nodes in $\xi$. If the number of nodes is zero, we have found $\omega_0^2$, and if it is not, we repeat the procedure for a lower initial value of $\omega^2$ (see Eq. \ref{eq_oms}).

In order to test our method, we construct numerical Lane-Emden Polytropes. We check that the polytropes satisfy $\omega^2=0$ at $\Gamma_1=4/3$ in the Newtonian case and $\Gamma_1=4/3+\kappa \frac{2GM}{Rc^2}$ in the relativistic case, where $M$ is the mass of the star, R the radius, and $\kappa$ is a constant determined numerically \citep{chandrasekhar1964}. Fig. \ref{fig:nconv_inst} shows the accuracy of these relations for increasing numerical resolution of the polytropes. We also verify $\xi_0 \propto r$ for these values of $\Gamma_1$, a condition that should hold for all polytropes.

Once an unstable model is found in the stellar code (e.g. Fig. \ref{fig:insta_hoshi}), the calculation is mapped to the hydrodynamical code (HYDnuc, Sec. \ref{methods_HYDnuc}). There, the star will either begin to collapse or stabilize due to nuclear burning. It is important to note that the stability analysis considers the stellar structure at a single moment, and cannot account for the energy generated by nuclear burning. 

Thus, from the start of the HYDnuc calculation, there is a competition between the growing perturbation of the unstable model and nuclear energy generation. For lower mass models ($M \leq 2.7  \times 10^4 \;\msun$), energy generation can sometimes stabilize the star. Fig. \ref{fig:insta_hydro} shows the stability analysis applied to a HYDnuc model which stabilizes (M = 2 $\times 10^4$ $\msun$, first instability). The stability analysis assumes zero velocity, so we cannot rely on it too heavily in a dynamical scenario, but it can be illustrative. The model is initially unstable and begins to contract; the star continues to contract, but the stability analysis fluctuates between 'stability' and 'instability'. Here, 'instability' and 'stability' refer to whether the contraction is growing exponentially or not. In order for the star to become stable in the usual sense, nuclear burning must increase to counteract the perturbation, which occurs around $10^5$ s. Afterwards, the temperature decreases and reaches a new equilibrium which is higher than that of the initial model.

For the high mass models (and later time low mass models), the perturbation overcomes the energy generation, and the SMS moves onto a collapsing phase that triggers rapid alpha capture burning and may lead to a GRSN. 

If the result of the hydrodynamical code is that the star stabilizes (as in Fig. \ref{fig:insta_hydro}), we map the next instability in the stellar code to HYDnuc and repeat until a model either explodes, pulsates, or collapses. In the case of e.g. 2 $\times 10^4$ $\msun$, the first such model has burned most of its helium (Table \ref{tab:models}).

\subsection{Hydrodynamics}
\label{methods_HYDnuc}

HYDnuc is a 1D GR implicit Lagrangian code \citep{takahashi2016,takahashi2018,takahashi2019,yoshida2019,nagele2020,nagele2021} based on the nuRADHYD code \citep{yamada1997,sumiyoshi2005,nagele2021} which includes a Boltzmann solver for neutrino transport not present in HYDnuc. HYDnuc uses a Roe-type approximate linearized Riemann solver. The independent variables are density, velocity, internal energy, entropy, electron fraction, radius, baryonic mass, enthalpy, and the two components of the Misner Sharp metric \citep{misner1964}. Internal energy changes primarily via energy generation from a nuclear network or cooling from thermal neutrino reactions. The equation of state is the same as in HOSHI.

In order to transport a model from HOSHI to HYDnuc, we first determine the radial values of the HYDnuc mesh according to the frequency function in Appendix B of \citet{takahashi2019}. This method is used to provide appropriate resolution to both the core and envelope of the SMS. Other variables are then inferred using linear radial interpolation of the log values of those variables, which we have found is the most accurate method of ensuring the fidelity (as a function of enclosed mass) of the remapping. Care must be taken at this step so as not to introduce unphysical perturbations. This is the same reason that we introduced the relativistic energy density to HOSHI (Sec. \ref{methods_HOSHI}). However, an unavoidable coordinate perturbation due to numerical error will always be present when changing mesh definitions, and this is a potential source of error in our simulation.

We continue the calculations for the explosions until there are convergence problems due to poor spatial resolution in the outer regions of the star. This typically occurs when the shock reaches $10^{15-16}$ cm. For the pulsations, a similar stopping condition is reached, but in this case we excise the ejected material from the simulation and verify that the remaining material becomes hydrostatic. For the models which collapse, we continue the calculation to a central temperature of 1 MeV, at which point neutrino heating begins to play a role.  

In order to set the numerical parameters for HYDnuc, we perform several numerical convergence tests using the the explosion energy, which is defined as the total energy when the shock reaches the stellar surface. Fig. \ref{fig:exp_nconv} shows the explosion energy as a function of mesh point number, total isotope number, and $\mathcal{V}$, the limit on the maximum variation of the independent variables 
\begin{equation}
    \mathcal{V}(k) > \rm{max}_{i,j} \; \Bigg|  \frac{x_{i}(j,k-1)}{x_{i}(j,k)} \Bigg|^{\pm 1}
    \label{eq_varlim}
\end{equation}
where $x_i$ is one of the independent variables of HYDnuc \citep{yamada1997}, j is the mesh point number, and k is the time-step. If the limit in Eq. \ref{eq_varlim} is violated for a time-step k, that time-step is repeated with a reduced value of dt. Thus, a smaller $\mathcal{V}$ will require more time-steps which increases the time resolution of the simulation, making it more physically accurate. The simulations with greater resolution tend to reach slightly higher temperature, which correlates to an increase in $E_{\rm exp}$. In this paper, we use $\mathcal{V}=10^{-5}$, a 61 isotope network, and 767 mesh points. The 61 isotope network includes the p-p chain, CNO and hot CNO cycles, triple alpha, detailed alpha process until silicon, a more basic alpha process up to nickel, and photodissociation of heavy elements.

The explosion energy depends on $\mathcal{V}$ and mesh point number in straightforward ways, but the dependence on isotope number requires explanation. The main driver of the explosion is alpha capture reactions, but not all of these reactions proceed at the same rate. In particular, the carbon alpha capture rate is lower than the reactions involving $^{20}$Ne and $^{24}$Mg; we use 1.5 $ \times $ the rate of \citet{caughlan1988}, but have verified that the explosion energy depends very weakly on this reaction rate. $\mathcal{O}(10\%)$ of the mass of the star is carbon, and very little of this is burned via carbon alpha capture on the timescale of the explosion. However, if nucleons are present, catalysis enhances the carbon alpha capture rate with 
\begin{equation}
    \rm ^{12}C(p,\gamma) ^{13}N, \;\; ^{13}N(\alpha,p) ^{16}O.
\end{equation}

In the networks with isotope number less than 61 (Fig. \ref{fig:cx_t}, left panel), a reservoir of free nucleons is built up during the explosion (Appendix B), and these then serve as the catalyst for carbon burning. In the networks with higher isotope numbers, however, the nucleons are absorbed in reactions such as 
\begin{equation}
    \rm ^{24}Mg(p,\gamma) ^{25}Al.
\end{equation}
Indeed, aluminium is of particular importance (Fig. \ref{fig:cx_t}, right panel) because the inclusion of its isotopes is the only difference between the 58 isotope network and the 61 isotope network, and from Fig. \ref{fig:exp_nconv}, we can see that this is the isotope number where the explosion energy converges. Appendix B contains a steady state calculation verifying this explanation. 

Usually, the inclusion of a larger network increases the nuclear energy generation, but in this case, a threshold number of elements is required to properly follow the nucleonic reactions. \citet{nagele2020} used a 49 isotope network and \citet{chen2014} used a 19 isotope network, so it is possible that the explosion energies found in those works are overestimated. In this work, the maximum explosion energy is a factor of 3-4 smaller than in the previous works. One reason for this is that the lower mass means there is less fusion material, but the effect of the nuclear network also plays a role. For a more detailed discussion of how this work compares to previous ones, see Sec. \ref{discussion}.

\begin{table*}
	\centering
	\caption{Summary table for all models. The columns are total mass, outcome of HYDnuc, mass of the isentropic core, central helium mass fraction at the start of HYDnuc, change in helium mass fraction, explosion energy, maximum central temperature, and maximum velocity of the outermost mesh point, denoted $v_R$.}
	\label{tab:models}
	\begin{tabular}{|c|l|l|l|l|l|l|l|} 
		
    		M [$10^4$ $\msun$] & Outcome & $M_{\rm core}$ [$\msun$] & $X_{\rm c}(^4$He) & $\Delta \; X_{\rm c}(^4$He) & $E_{\rm exp}$ [ergs] & max $T_{ \rm c} $ [K]& max $v_R$/c  \\
    		\hline
2 & Collapse & 10926 & 1.37e-3 &  ---  &  ---  &  ---  &  ---  \\ 
2.1 & Collapse & 11368 & 2.23e-4 &  ---  &  ---  &  ---  &  ---  \\ 
2.2 & Collapse & 11729 & 1.22e-4 &  ---  &  ---  &  ---  &  ---  \\ 
2.3 & Collapse & 12595 & 3.17e-2 &  ---  &  ---  &  ---  &  ---  \\ 
2.4 & Collapse & 13180 & 3.44e-18 &  ---  &  ---  &  ---  &  ---  \\ 
2.5 & Collapse & 13798 & 2.69e-3 &  ---  &  ---  &  ---  &  ---  \\ 
2.6 & Pulsation & 14772 & 0.104 & 0.104 & 4.32e53 & 7.58e8 & 0.032 \\ 
2.7 & Pulsation & 14964 & 0.222 & 0.147 & 4.70e52 & 6.62e8 & 0.021 \\ 
2.8 & Collapse & 15596 & 0.713 &  ---  &  ---  &  ---  &  ---  \\ 
2.9 & Pulsation & 16183 & 0.589 & 0.153 & 7.56e53 & 7.33e8 & 0.041 \\ 
2.95 & Explosion & 16504 & 0.599 & 0.168 & 1.23e54 & 7.69e8 & 0.046 \\ 
3 & Explosion & 16817 & 0.652 & 0.152 & 1.43e54 & 8.06e8 & 0.048 \\ 
3.05 & Collapse & 17144 & 0.734 &  ---  &  ---  &  ---  &  ---  \\ 
3.1 & Collapse & 17516 & 0.794 &  ---  &  ---  &  ---  &  ---  \\ 
3.15 & Collapse & 17793 & 0.815 &  ---  &  ---  &  ---  &  ---  \\ 
3.2 & Collapse & 18091 & 0.815 &  ---  &  ---  &  ---  &  ---  \\ 
3.3 & Collapse & 18888 & 1.000 &  ---  &  ---  &  ---  &  ---  \\ 
3.4 & Collapse & 19460 & 1.000 &  ---  &  ---  &  ---  &  ---  \\ 
3.5 & Collapse & 19933 & 0.950 &  ---  &  ---  &  ---  &  ---  \\ 
4 & Collapse & 23891 & 0.960 &  ---  &  ---  &  ---  &  ---  \\ 

		\hline
	\end{tabular}
\end{table*}

\begin{table*}
	\centering
	\caption{Mass ejecta by isotope for the explosions and the pulsations. Except for the first column which is consistent with Table \ref{tab:models}, values are recorded in units of $\msun$. Yield tables for the explosions are available online.}
	\label{tab:ejecta}
	\begin{tabular}{|c|l|l|l|l|l|l|l|l|l|} 
		
    		M [$10^4$ $\msun$] &M$_{\rm ej}$ & M($^1$H)  & M($^4$He) & M($^{12}$C) & M($^{16}$O) & M($^{20}$Ne) & M($^{24}$Mg) & M($^{28}$Si) & M($^{32}$S)  \\
    		\hline
2.6 & 2808 & 1877 & 974 & < 0.1 & < 0.1 & < 0.1 & < 0.1 & < 0.1 & < 0.1 \\
2.7 & 2299 & 1584 & 759 & < 0.1 & < 0.1 & < 0.1 & < 0.1 & < 0.1 & < 0.1 \\
2.9 & 2078 & 1465 & 651 & < 0.1 & < 0.1 & < 0.1 & < 0.1 & < 0.1 & < 0.1 \\
2.95 & 29500 & 5441 & 16946 & 3006 & 2505 & 481 & 812 & 306 & 1.2 \\
3 & 30000 & 5537 & 18077 & 2986 & 1829 & 367 & 702 & 497 & 5.1 \\

		\hline
	\end{tabular}
\end{table*}

\begin{figure}
    \centering
    \includegraphics[width=\columnwidth]{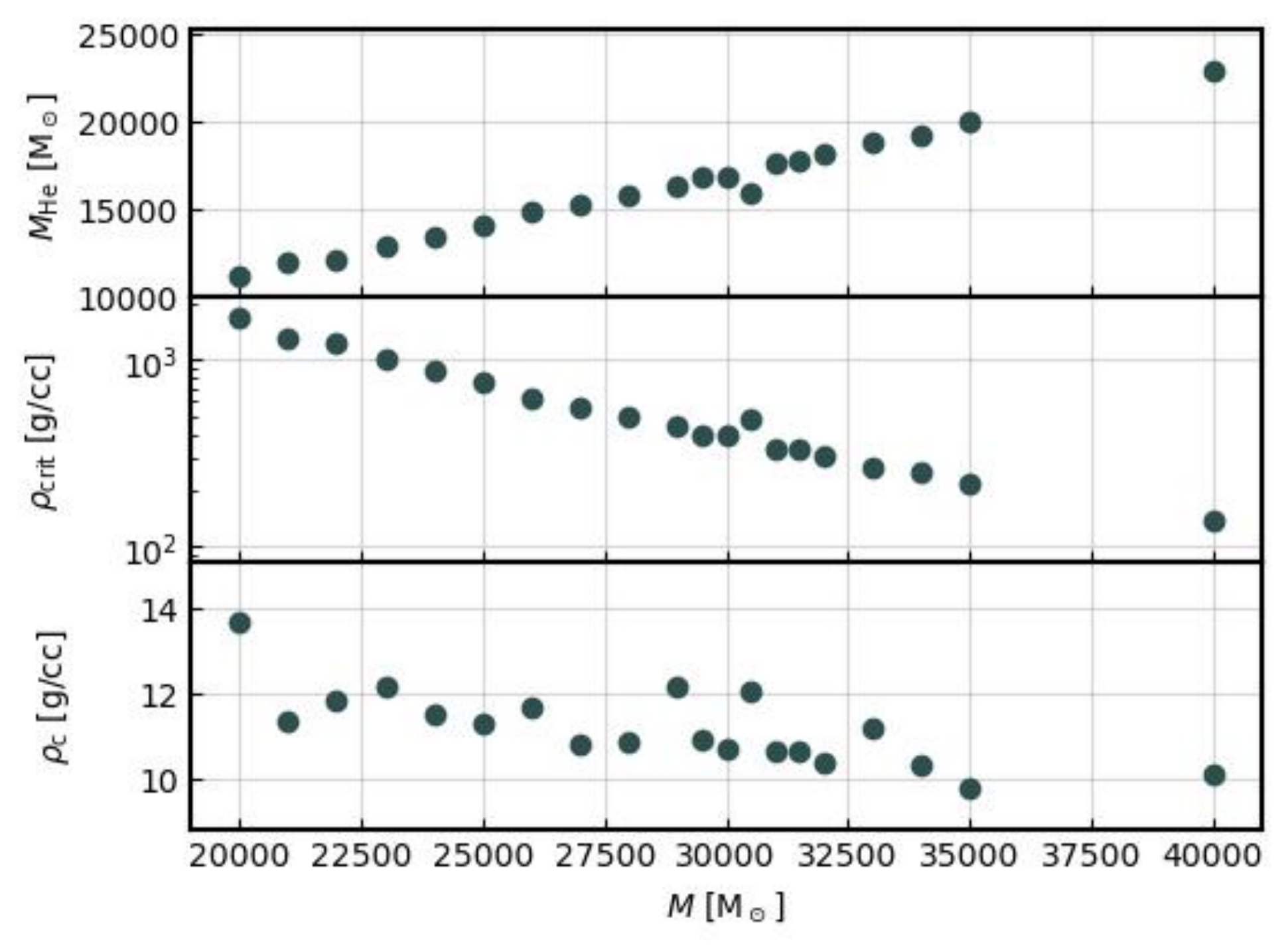}

    \caption{Illustration of the polytropic criterion for the models in this paper. Upper panel --- mass of the helium core from HOSHI, determined by the mass outside which X($^1$H) $>$ 1e-5. Middle panel --- critical density (Sec. \ref{results_comp}) of a pure helium core with the mass from the top panel. Lower panel --- central density of the HOSHI models at the maximum central helium mass fraction. A comparison of the middle and lower panels shows that these models are not yet unstable according to the polytropic criterion.}
    \label{fig:polytropic}
\end{figure}

\begin{figure}
    \centering
    \includegraphics[width=\columnwidth]{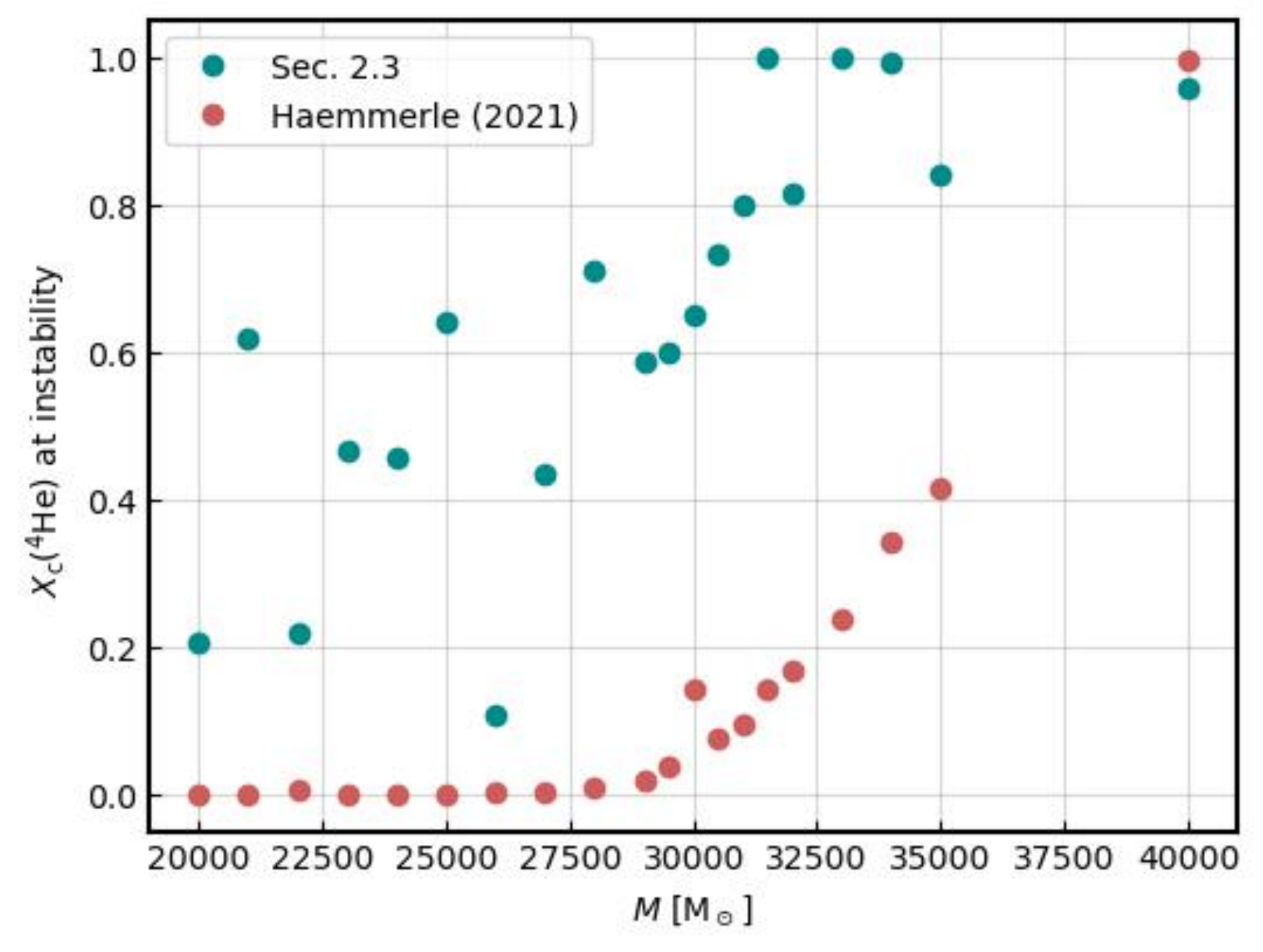}

    \caption{Comparison of the first instability reached by two methods, Sec. \ref{methods_stab} and that of \citet{haemmerle2020}. Note that the values in this figure differ from Table \ref{tab:models} for lower masses because the first instability does not always collapse in HYDnuc (Sec. \ref{methods_stab}).} 
    \label{fig:haemmerle}
\end{figure}

\section{Results}
\label{results}

The primary advantage of this paper is the GR stability analysis, so before discussing applications, we will compare this analysis to two previous methods.

\begin{figure*}
    \centering
    \includegraphics[width=2\columnwidth]{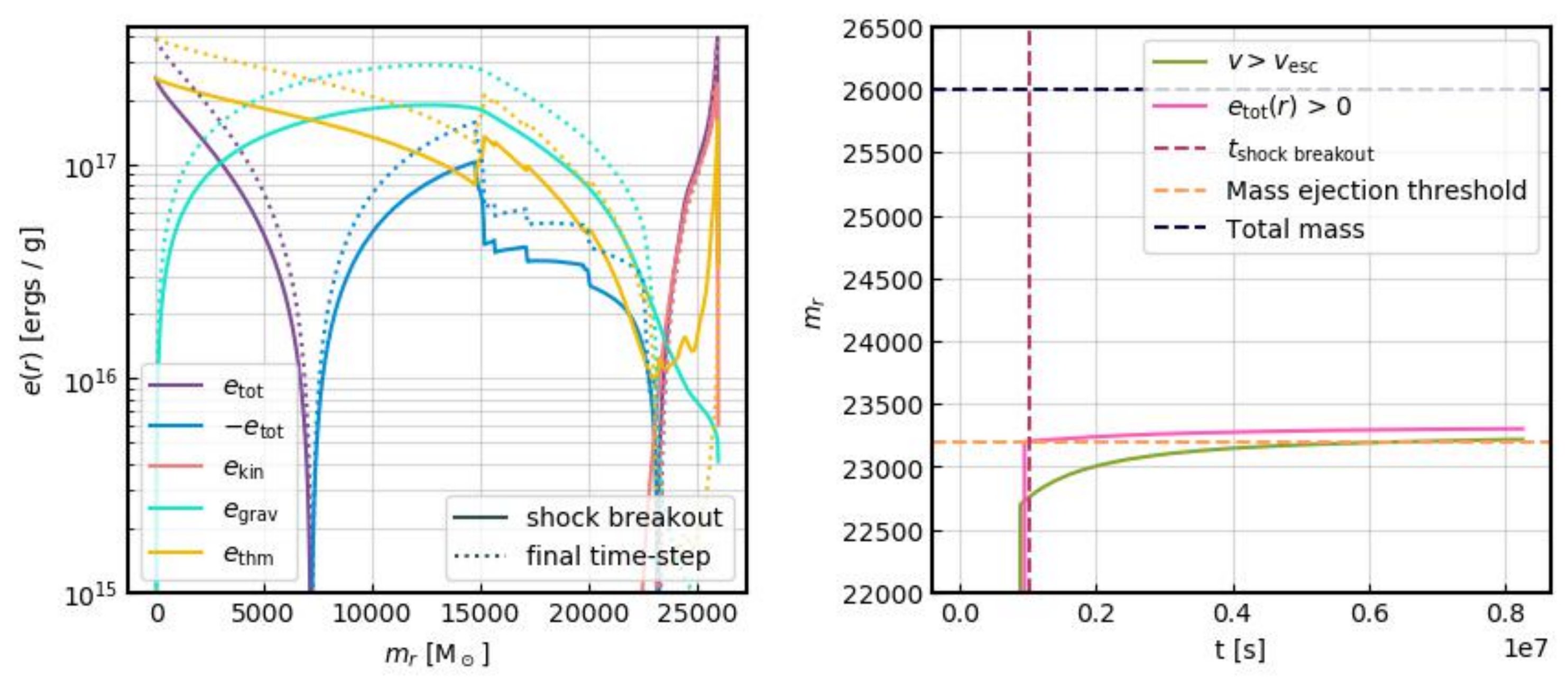}

    \caption{Left panel --- local energy as a function of mass coordinate for the 2.6 $\times 10^4$ $\msun$ model (pulsation). The ejection criterion is $e_{\rm tot}>0$ for all $m_r > \rm M - M_{\rm ej}$ at the time of shock breakout. Right panel --- illustration of the stability of the ejection criterion. We measure the ejected mass at shock breakout and this quantity is roughly consistent with the post Newtonian escape velocity criterion near the end of the simulation. .}
    \label{fig:E_pulse}
\end{figure*}

\subsection{Comparison to previous works}
\label{results_comp}

First, and most common in the literature \citep[e.g.][]{fuller1986,umeda2016,woods2017} is the polytropic criterion. SMSs have high entropy and are supported mostly by radiation pressure and this invites analytic approximations. In particular, it is often assumed that the SMS core is very nearly an $n=3$ polytrope. 

The explosion in the next section involves explosive helium burning, so we will calculate the instability condition for a polytrope consisting of pure helium, as a function of the mass of the helium core ($M_{\rm He}$) which is taken from the stellar evolution calculation. Once the mass of the helium core is known, the radiation entropy may be approximated as $s_r \propto M_{\rm He}^{1/2}$ \citep{shapiro1983}. From here, the polytropic constant is determined
\begin{equation}
    K=\frac{a}{3} \bigg( \frac{3s_r}{4m_pa} \bigg)^{4/3}.
\end{equation}
Next, we calculate the outer radius at which the star will be unstable, $R_{\rm crit}$ by setting the SMS $\Gamma$ equal to the general relativistic $\Gamma_1$, 
\begin{equation}
     \frac{4}{3} + \frac{\beta}{6} + \mathcal{O}(\beta^2) = \frac{4}{3} + \frac{2GM_{\rm He}\kappa}{R_{\rm crit} c^2}
    \label{eq:rcrit}
\end{equation}
where $\beta \approx 4.3/\mu (M/\msun)^{-1/2}$ is the ratio of the gas pressure to total pressure, and $\kappa = 2.249$ for $n=3$ (this form of $\kappa$ differs from \citet{chandrasekhar1964} by a factor of 2). Eq. \ref{eq:rcrit} can be solved for $R_{\rm crit}$ as a function of $M_{\rm He}$, so that we finally arrive at an expression for the critical density \citep{shapiro1983}
\begin{equation}
    \rho_{\rm crit} = \Bigg[\frac{R_{\rm crit}}{\xi_1} \bigg( \frac{K}{\pi G}\bigg)^{-1/2} \Bigg]^{-3} \propto M_{\rm He} ^ {-7/2}
\end{equation}
as a function of $M_{\rm He}$, where $\xi_1 = 6.897$ is determined numerically.

Thus, we have an expression for the critical density of a purely helium SMS core as a function of $M_{\rm He}$. By construction, the SMS models in this paper are near to this point according to the GR stability analysis from Sec. \ref{methods_stab}. Fig. \ref{fig:polytropic} shows that, for these models, the central density when the star is pure helium is more than an order of magnitude below the critical density. So, the polytropic criterion underestimates the GR instability in comparison to Sec. \ref{methods_stab}.

Next we compare our method to the results of \citet{haemmerle2020}, who also evaluate Eq. \ref{eq_GR}, though they make the simplifying assumption $\xi \propto r$. This assumption is valid for polytropic stars and possibly for higher mass hydrogen SMSs, but for our models, the perturbation is not always proportional to the radius (see Appendix A). 

Fig. \ref{fig:haemmerle} shows the helium mass fraction at the first instability --- that is, the first model in the HOSHI calculation which is unstable --- determined by either method. The central helium mass faction for an explosion or pulsation is roughly 0.1 < $X_{\rm c}(^4$He) < 0.7, c.f. Table \ref{tab:models}. Our method uses a necessary condition for instability while the method of \citet{haemmerle2020} uses a sufficient condition. This means that there are models which would be stable according to the method of \citet{haemmerle2020}, but not according to our method. Furthermore, any model which is unstable according to the method of \citet{haemmerle2020} will also be unstable according to our method. Thus, our method will find an instability earlier in the evolutionary calculation than the method of \citet{haemmerle2020}. The longer the time before the instability in the evolutionary calculation, the higher the mass range for the explosion, because lower mass models which would explode if the instability occurred earlier instead burn all of their helium ($M<3 \times 10^4$ in Fig. \ref{fig:haemmerle}). This means that the explosion would occur at larger mass, if we were to use the method of \citet{haemmerle2020}, although that being said, the explosion mass range found by their method would still be significantly smaller than the case of $M\sim 5 \times 10^4 \; \msun$ discussed in \citet{chen2014,nagele2020}.

\begin{figure*}
    \centering
    \includegraphics[width=2\columnwidth]{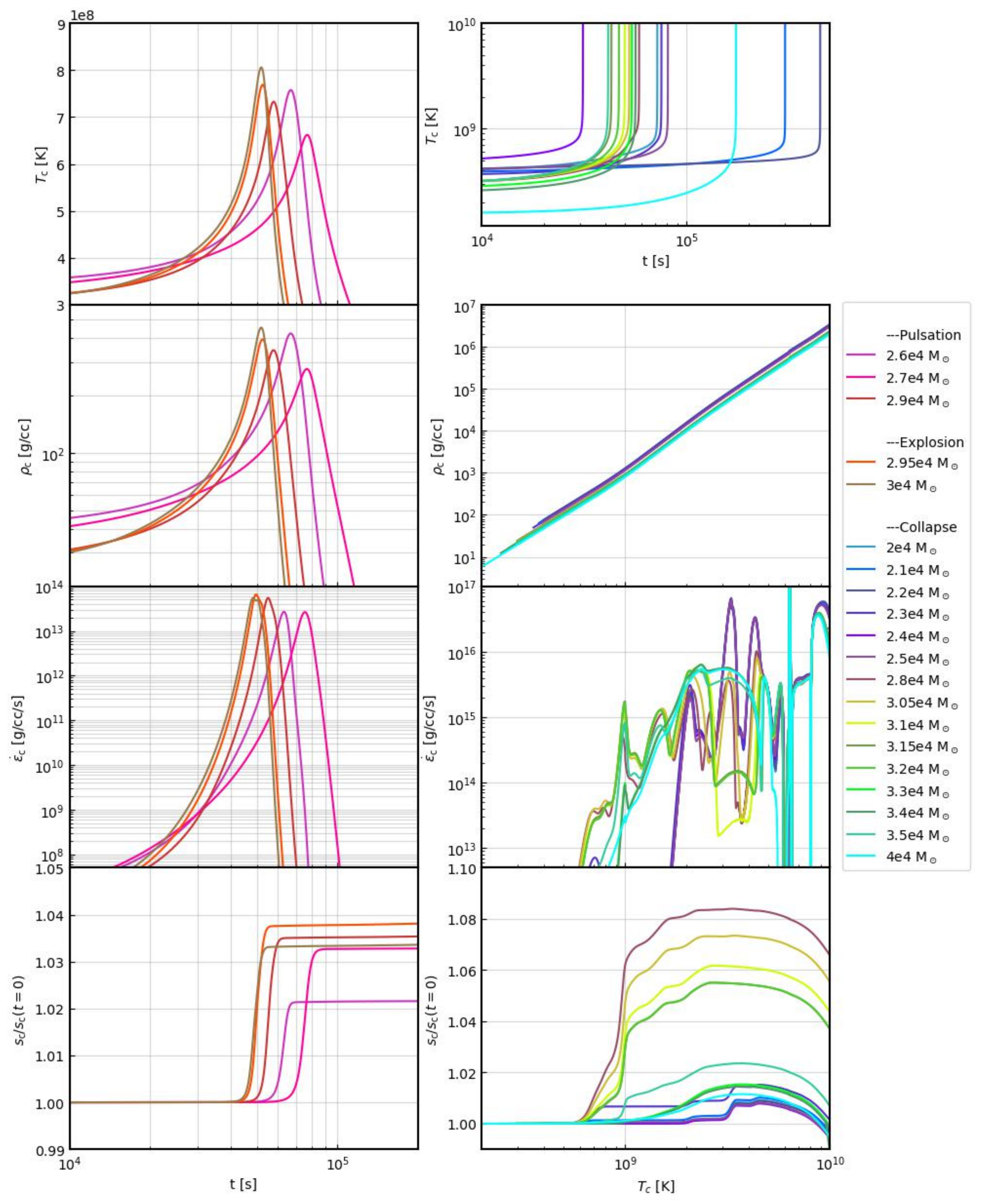}

    \caption{Time evolution of central temperature (1st panel), density (2nd panel), rate of change of specific internal energy (3rd panel) and entropy relative to the initial value (4th panel). The legend groups models by outcome, whereas colors vary with mass. The left column shows the exploding and pulsating models as a function of time. The right column shows the temperature as a function of time, while the other three panels are functions of temperature. }
    \label{fig:central}
\end{figure*}

\begin{figure*}
    \centering
    \includegraphics[width=2\columnwidth]{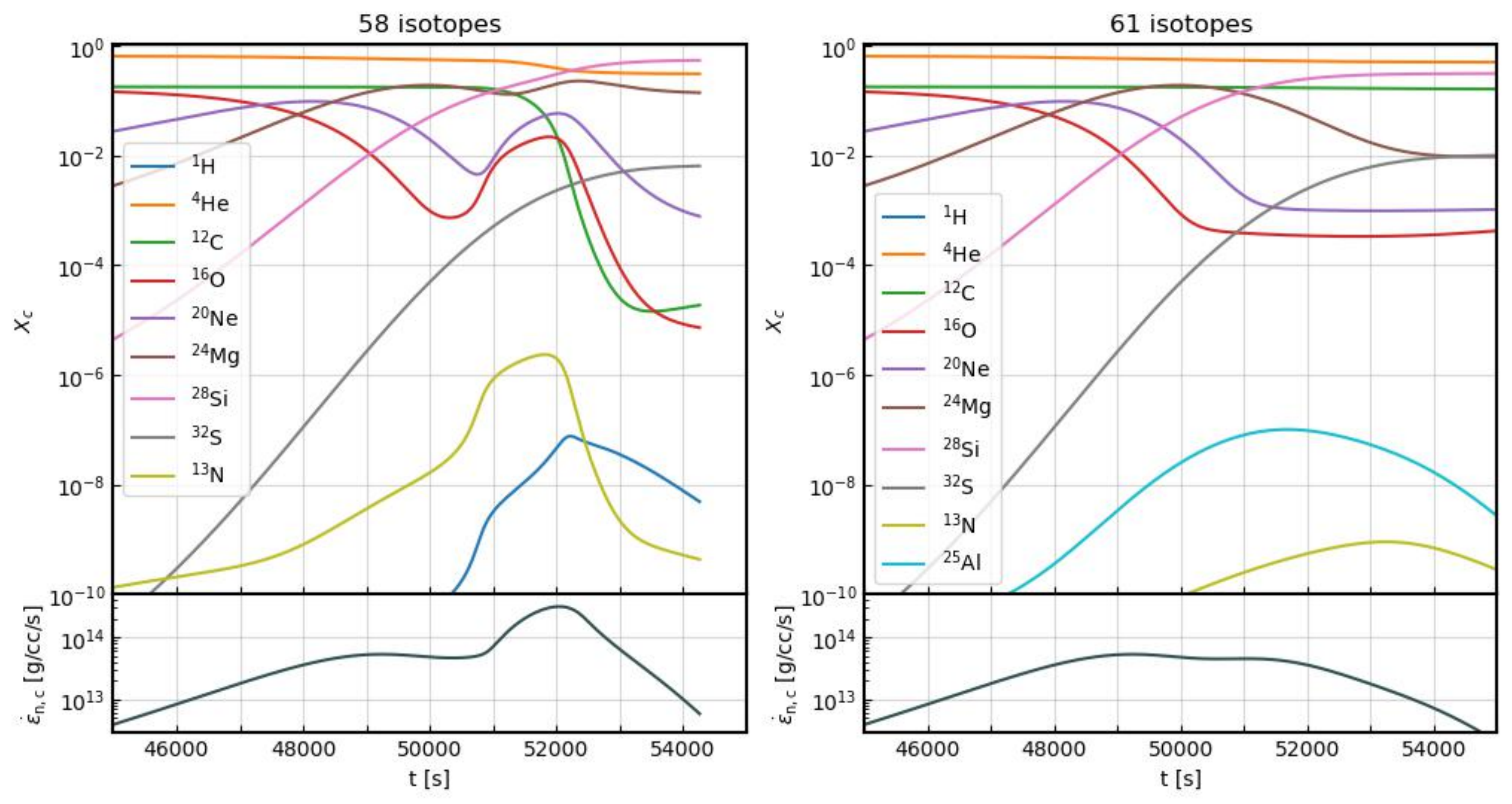}

    \caption{Upper panel --- comparison of central isotope mass fractions for the M = 3 $\times 10^4$ $\msun$ model with a 58 isotope model (left) and a 61 isotope model (right). The high mass fraction of protons and $^{13}$N in the left panel facilitates artificially high carbon burning. Lower panel ---  time evolution of central $\dot{\epsilon}$.}
    \label{fig:cx_t}
\end{figure*}

\begin{figure}
    \centering
    \includegraphics[width=\columnwidth]{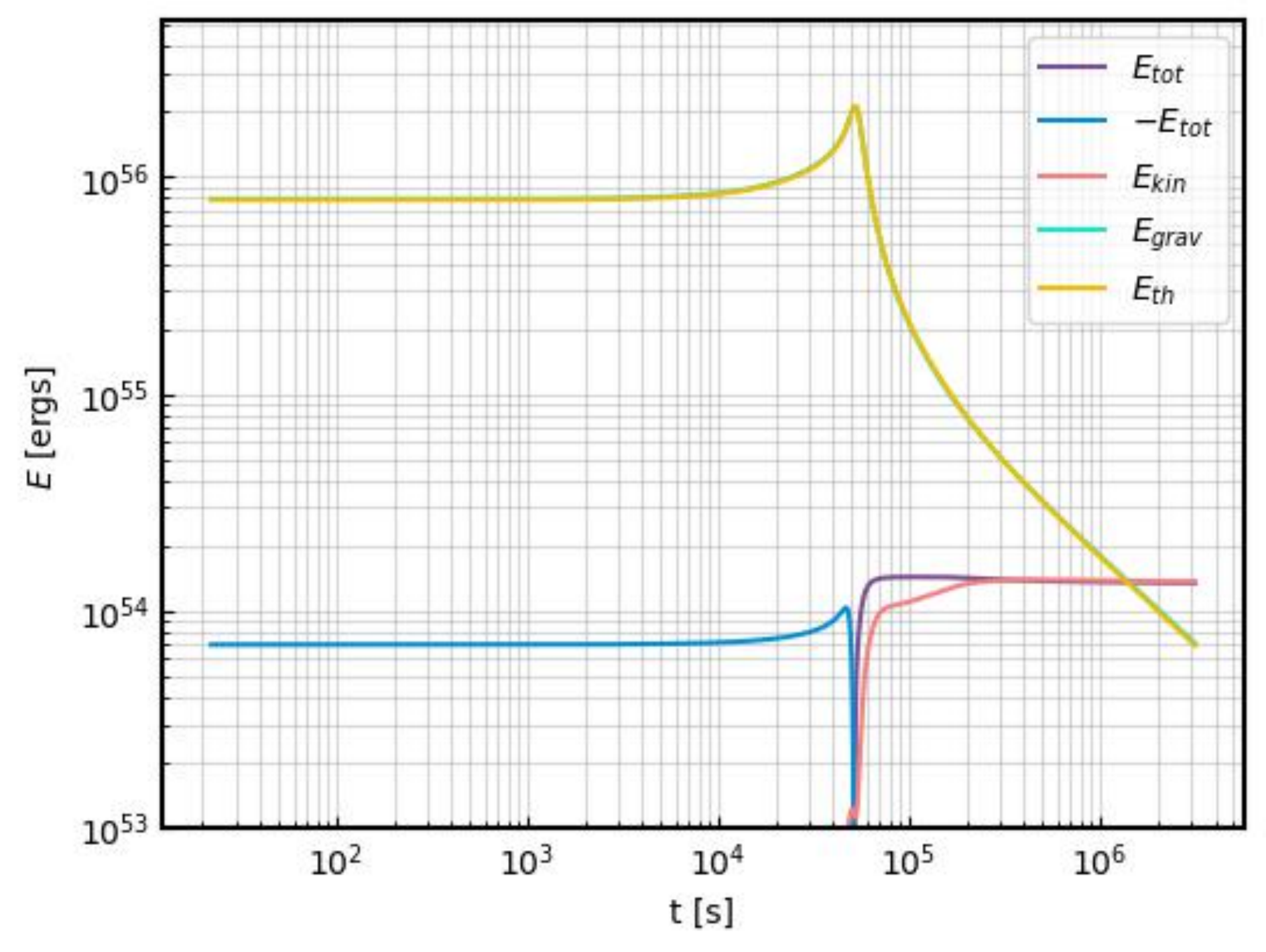}

    \caption{Total energy, kinetic energy, gravitational energy, and thermal energy (as defined in \citet{nagele2020}) in HYDnuc as a function of time for the explosion of the 3 $\times 10^4$ $\msun$ model. Unless otherwise specified, subsequent figure will show the 3 $\times 10^4$ $\msun$ model.}
    \label{fig:E_explo}
\end{figure}

\begin{figure*}
    \centering
    \includegraphics[width=2\columnwidth]{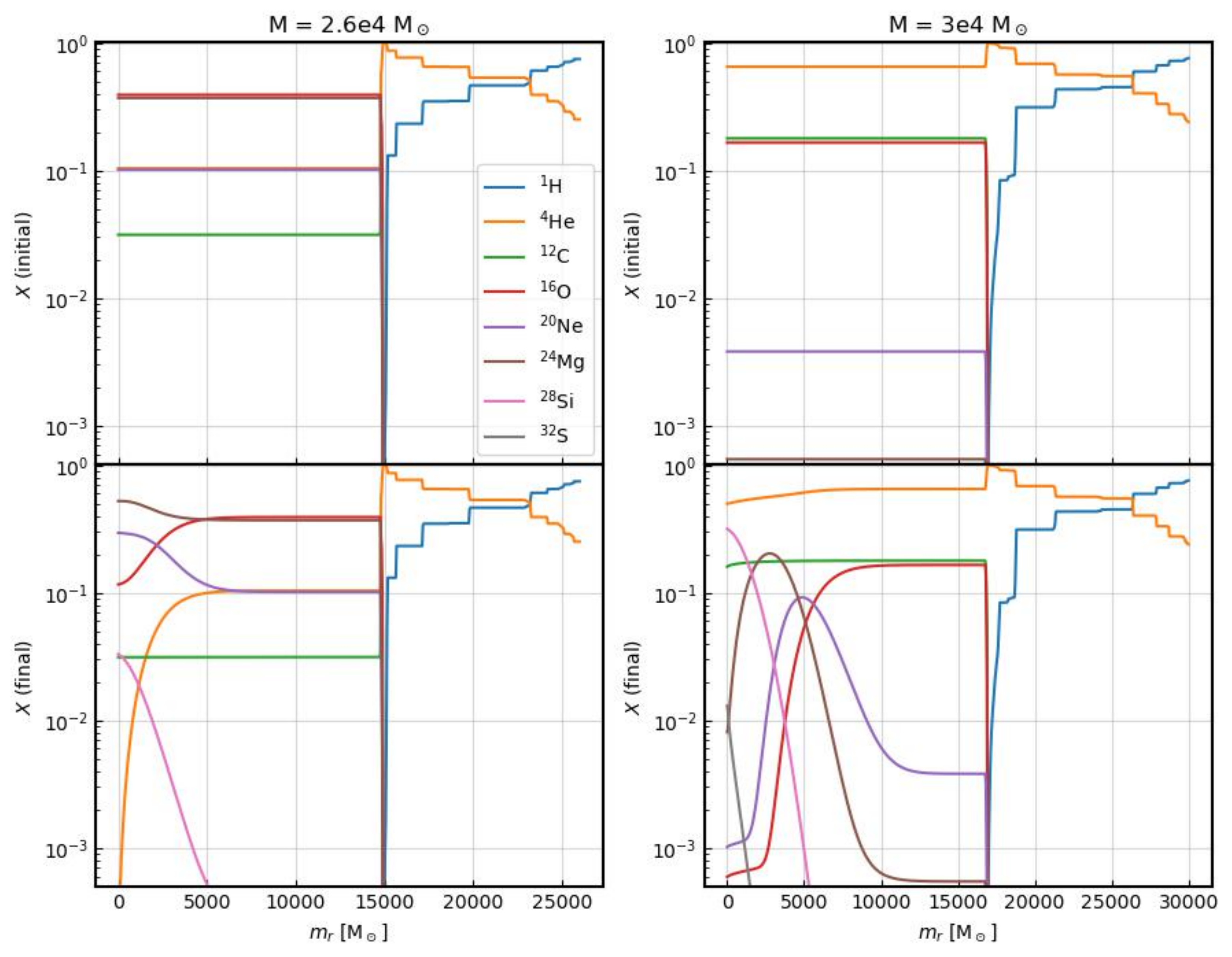}

    \caption{Initial (upper panel) and final (lower panel) isotope mass fractions as a function of the mass coordinate for the pulsation, M = 2.6 $\times 10^4$ $\msun$ (left) and the explosion, M = 3 $\times 10^4$ $\msun$ (right). }
    \label{fig:cx_mr}
\end{figure*}

\begin{figure*}
    \centering
    \includegraphics[width=2\columnwidth]{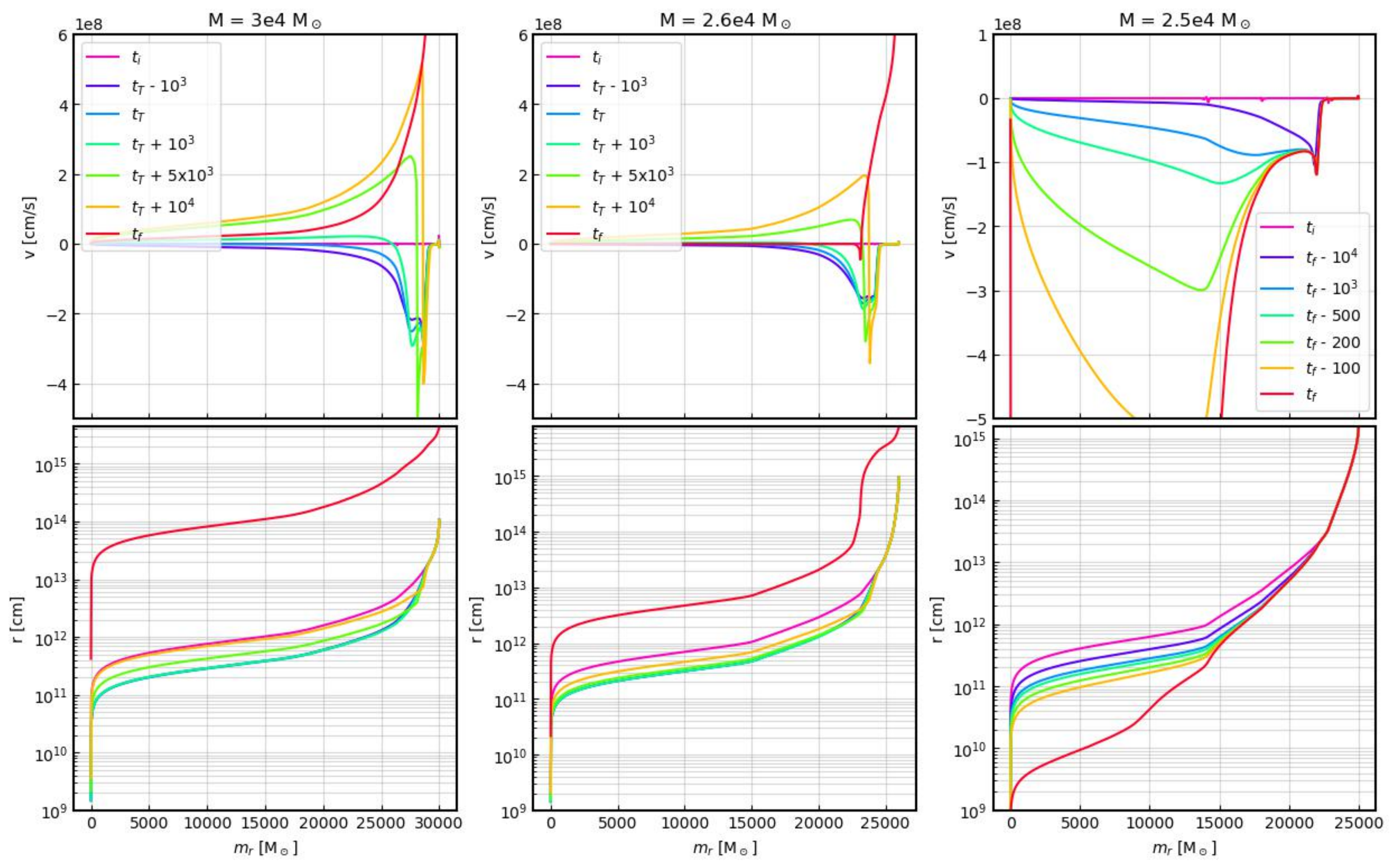}

    \caption{Upper panel --- velocity snapshots at several time-steps ($t_i$ is the first time-step in HYDnuc, $t_f$ the last, and $t_T$ the time-step with maximum central temperature, see Table \ref{tab:models}) for an exploding model (left), a pulsating model (middle) and a collapsing model (right). Lower panel --- same as upper but for radius.}
    \label{fig:v}
\end{figure*}

\begin{figure}
    \centering
    \includegraphics[width=1\columnwidth]{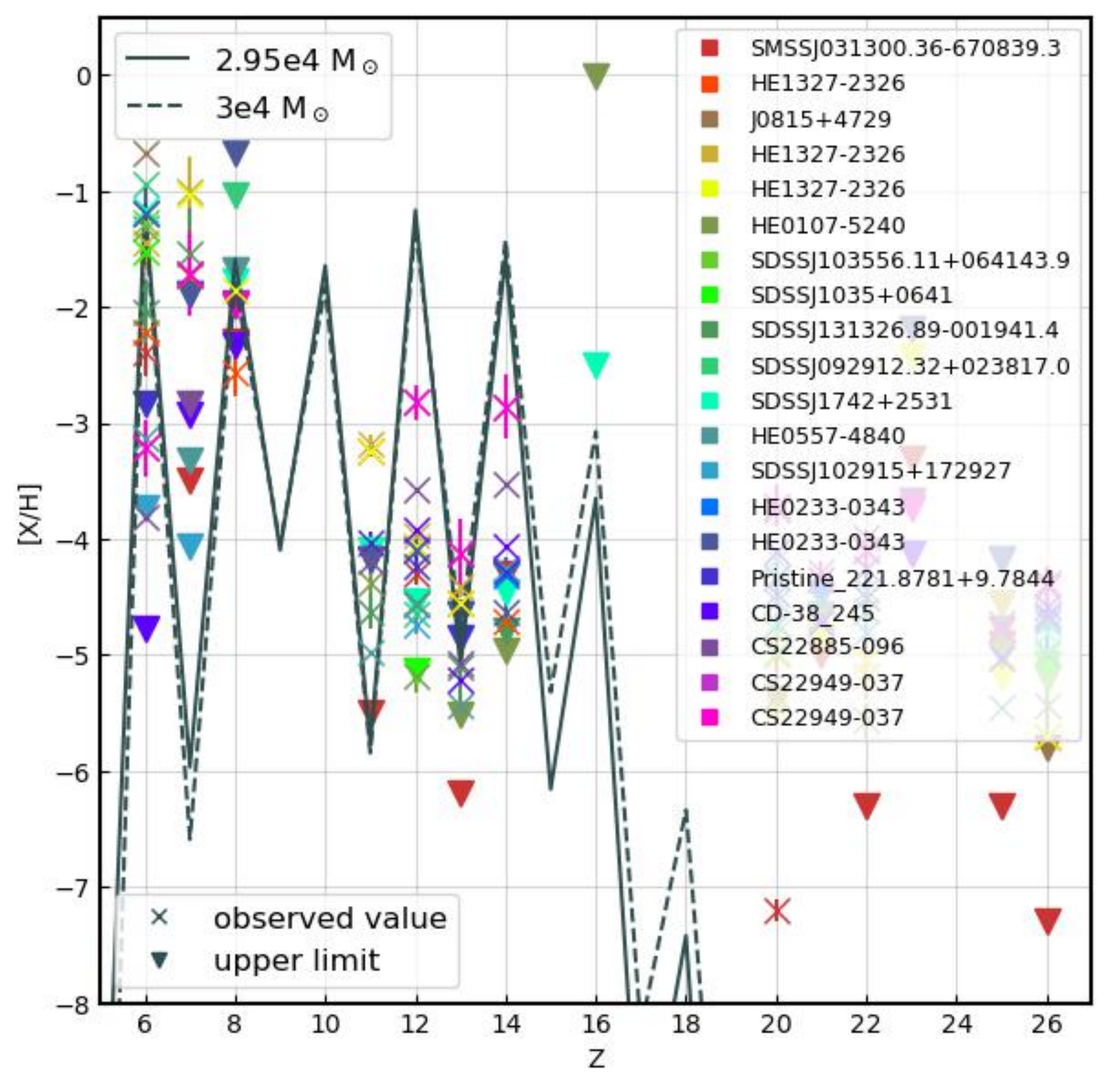}

    \caption{Explosion yield (lines) compared with observed metal poor stars (crosses are observations, triangles upper limits). The line shows the fraction relative to a minimum mass of hydrogen with which it would have to mix \citet{magg2020}, and can thus be regarded as an upper limit. }
    \label{fig:XH}
\end{figure}

\subsection{Application to GRSNe}
\label{results_explo}

\subsubsection{HYDnuc explosion}

As hinted at in previous sections, we find two GRSNe as well as several pulsations (Table \ref{tab:models}). The pulsations are less energetic events which eject a portion of the envelope (Table \ref{tab:ejecta}). The relationship between the GRSNe and these pulsations is likely similar to the relationship between pair instability supernovae and the pulsational pair instability process \citep{woosley2017}. In \citet{nagele2020}, we did not find any pulsations because by the time we switched to HYDnuc (at the end of the HOSHI calculation), the star was extremely unstable, meaning it could only collapse or, in one somewhat unique case, explode. The GR stability analysis allows us to find stars which have only just become unstable, and thus can achieve multiple outcomes (stabilize, explode, pulsate, collapse) when ported to HYDnuc. The widening of the mass window for the explosion and the discovery of pulsations are thus natural consequences of utilizing the GR stability analysis.

For the pulsations, we determine the ejecta mass in Table \ref{tab:models} using the local energy, e(r) (Fig. \ref{fig:E_pulse}, left panel), which is the integrand of the global energies defined in \citet{nagele2020}. We measure this quantity at shock breakout (right panel), when some of the energy is still in the form of thermal energy. The energy evolution after shock breakout may not be completely accurate because the energy is evolved using the entropy equation, and energy conservation is not guaranteed, especially in extremely low density regions where the accuracy of the EOS suffers. We confirm that the escape velocity criterion (right panel) converges roughly to this same value, validating the use of this ejection criterion. Note that $2.9 \times 10^4$ $\msun$ is a marginal case. Although it is not an explosion, it does not have the steady behaviour of Fig. \ref{fig:E_pulse} after shock breakout, and we expect the value in Table \ref{tab:ejecta} to underestimate the ejecta mass for this case.

The mass range for the explosion follows from straightforward considerations. Models which experience the instability before helium burning has reduced the binding energy of the star cannot explode. On the other hand, models with helium mass fractions less than ten percent also do not explode or pulsate because they lack fuel for alpha capture reactions sufficient to halt the collapse.

The pulsating models all have lower mass than the exploding models and the most massive model being the most energetic before a sharp drop-off is reminiscent of Fig. 6 of \citet{nagele2020}. This is a likely characteristic of the GRSNe even if the mass range found by the current analysis is not completely correct.

Fig. \ref{fig:central} shows the behavior of the central temperature, baryonic density, nuclear energy generation rate, and entropy as a function of time. For the exploding and pulsating models (left column), temperature and density increase and decrease smoothly, as was the case for the GRSNe in \citet{chen2014,nagele2020}, and the nuclear energy generation rate is also smooth. The models which collapse (right column) also show relatively smooth increases in temperature and density, but the energy generation rate and change in entropy are more complicated, as the stars enter different phases of burning and photodissociation. Roughly speaking, the first three peaks in $\dot{\epsilon}_c$ (first peak at $10^9$ K) correspond to carbon burning, sulfur burning, and calcium to iron peak element burning. Stars with a more evolved core, such as 2.4 $\times $ $10^4$ $\msun$ do not have much carbon remaining, and the first peak corresponds to neon burning. At the other extreme, stars with reserves of hydrogen, such as 4 $\times $ $10^4$ $\msun$ do not exhibit peaks because the presence of protons increases the number of possible reactions. After the core becomes iron/nickel, the next major reactions are assisted by free nucleons created by photodissociation (one peak), which is finally followed by photodissociation of nickel (one peak) and photodissociation of helium (two peaks). In reality, neutrino reaction would then begin to dominate, but HYDnuc does not include most of the relevant neutrino reactions.

For the explosion of 3 $\times 10^4$ $\msun$--- which we will use as an example in the figures--- Fig. \ref{fig:E_explo} shows the time evolution of the energy quantities, including the total energy which eventually determines the explosion energy (Table \ref{tab:models}). There are three phases, the initial contracting phases with $E_{\rm tot}<0$, the pre-shock breakout phase with $E_{\rm tot}>E_{\rm kin}>0$, and the post-shock breakout phase with $E_{\rm tot}=E_{\rm kin}>0$. Fig. \ref{fig:cx_mr} shows the isotope mass fraction as a function of mass coordinate for the initial (upper) and final (lower) time steps. The SMS cores are initially isentropic and shell hydrogen burning often occurs. The envelope is divided into many convective layers, the exact layout of which can alter the stability of the star \citep{nagele2020}. The final isotope distributions show that the majority of the nuclear burning takes place within the inner 5000 $\msun$ for the pulsating model and the inner 10000 $\msun$ for the exploding model. For the exploding model, the star is totally disrupted (e.g. Fig. \ref{fig:v}), so these elements will be ejected into the inter stellar medium (ISM) (Table \ref{tab:ejecta}).

After the explosive nuclear burning, the inwards velocity rapidly reverses and the shock propagates towards the surface of the SMS (Fig. \ref{fig:v}) with a typical velocity of a few percent the speed of light (Table \ref{tab:models}); shock breakout occurs on a timescale of $10^5$ s. The initial inwards velocity is largest in the envelope, and we emphasize that the GRSN involves the collapse of the entire star, not just the core. This is another reason why the analysis in Sec. \ref{methods_stab} is necessary, as an analysis of the core alone will not always capture the instability. In the exploding case (Fig. \ref{fig:v}, left), the final velocity is monotonically increasing, while in the pulsating case (Fig. \ref{fig:v}, central) the final velocity nears zero in most of the star, is slightly negative at the edge of the remnant, before rapidly increasing with the ejected material. In the bottom row of this figure, the corresponding discontinuity in radius is clearly visible. 

\subsubsection{Comparison to observed metal poor stars}

We find a wider explosion window than previous works, as well as pulsations which could later explode. However, the explosion window is still narrow compared to the feasible mass range of SMSs. This narrowness, in combination with the probable paucity of SMSs themselves, means that the likelihood of a GRSN having occurred in the Milky Way volume where its archaeological imprints might feasibly be observed is low. With that disclaimer in place, however, we will now compare the GRSN ejecta to observed metal poor stars. 

In this section, we will consider only the mono-enrichment scenario (see e.g. \citealt{hartwig2018a,hartwig2019}), so that the metals in the metal poor star have come exclusively from the GRSN ejecta and thus we ignore potential pollution from ISM accretion. we determine the minimum mass of ISM material with which the ejecta could mix via Eq. 2 of \citet{magg2020} where the explosion energy is taken from Table \ref{tab:models} and we use the fiducial value of the number density from \citet{magg2020}, $n_0 = 1$ cm$^{-3}$. We combine this mass with our ejecta and plot the inferred abundances relative to hydrogen, relative to solar (Fig. \ref{fig:XH}). Since the mixing mass is a lower limit, the abundance ratios are an upper limit. 

Also shown are the inferred abundances of the twenty stars in the SAGA database \citep{suda2008} with the lowest values of Fe / H: \citet{keller2014,ezzeddine2019,aguado2018,aoki2006,frebel2008,christlieb2004,bonifacio2018,bonifacio2015,frebel2015,caffau2016,norris2007,caffau2011,hansen2014,hansen2015,starkenburg2018,roederer2014a,roederer2014b}. We select these stars because the GRSN produces negligible amounts of iron. Fig. \ref{fig:XH} shows that our inferred yields do not match any of the observed stars. Even for the star of \citet{keller2014} which has strict upper limits on [Fe/H], our yield misses the observed value of calcium by several orders of magnitude. 

While we can seemingly rule out the mono-enrichment scenario for observed metal poor stars (which may not be reflective of the entire population of metal poor stars) in the vicinity of our galaxy, the multi-enrichment scenario is more challenging to rule out. The best we can do at current is to note that although many metal poor stars are carbon enriched, they are not generally silicon and magnesium enriched, which is evidence against an SMS being involved in the multi-enrichment scenario for metal poor stars. We further note that it would be nearly impossible to rule out GRSN enrichment near the galactic center, because of chemical dilution and observational difficulties.

\subsection{Application to SMS Collapse and neutrino emission}
\label{results_col}

We compare the neutrino light-curves of the lowest (2 $\times 10^4$ $\msun$) and highest (4 $\times 10^4$ $\msun$) mass models in this study with the results from our previous work \citep{nagele2021}. The nuRADHYD code is unchanged so the only difference is the amount of time the star spends in the evolutionary stage. Both stars have higher entropy than in our previous work because there is less time for neutrino cooling during the evolutionary stage, and even though the models in this work have different chemical compositions at the start of collapse, they will eventually undergo the same reactions, namely alpha capture until sulfur, then production of calcium through nickel, followed by photodisassociation, first into helium and then into nucleons.

In our previous paper, we identified that many physical quantities scaled with the entropy at fixed density. Thus, the stars in this paper having higher entropy would suggest that they should also have higher temperature, and neutrino luminosity. Both of these turn out to be true, but while the hydrodynamical quantities match the trends in Fig. 8 of \citet{nagele2021}, the neutrino quantities do not. Thus, the neutrino luminosity, and number flux are all increased relative to the previous work, but not by as much as we expected. Of particular interest, the total neutrino number increased by 13$\%$ for 2 $\times 10^4$ $\msun$ and 42$\%$ for 4 $\times 10^4$ $\msun$. Furthermore, although we would expect the average neutrino energy to decrease because of the higher entropy, they increase by small fractions, 2$\%$ for 2 $\times 10^4$ $\msun$ and 11$\%$ for 4 $\times 10^4$ $\msun$.

Although these both trend in the right direction regarding detection of the diffuse SMS neutrino background, they are not large enough increases to alter our previous conclusion that if SMSs collapse in this mass range, then the detection of this background is not feasible using current methods.

\section{Discussion}
\label{discussion}

Using the general relativistic stability analysis in Sec. \ref{methods_stab}, we can more accurately predict when a SMS will be unstable and will collapse explode or pulse. This is necessary because the timescale of the collapse is many orders of magnitude shorter than the evolutionary timescale, meaning that our stellar evolution code often misses the GR instabilities. For some models, the instability is countered by increased nuclear burning, but for masses around 3 $\times 10^4$ $\msun$, this is not the case. The 2.95 $\times 10^4$ $\msun$ and 3 $\times 10^4$ $\msun$ models explode in GRSNe, while several lower mass models pulsate and eject portions of their envelope. The final fate of the pulsating models is unclear, as they will reenter the evolutionary track with different properties, most notably the chemical composition of the core and total energy. If multiple pulsations were to occur, it could cause a collisional supernova.

In comparison to the GRSNe from \citet{chen2014,nagele2020}, our GRSNe have much lower explosion energies. There are a few differences between the current work and \citet{chen2014}. The most important of these is likely the timescale over which the explosion takes place. The PN approximation employed in \citet{chen2014} is extremely accurate (although see Fig. \ref{fig:HOSHI_r}), but it only includes the approximation to the hydrostatic terms in the equation of motion. In a dynamical scenario, such as the GRSN, the velocity becomes large, and the hydrodynamical PN terms should also be included to properly follow the dynamics. Because of this, the approach of \citet{chen2014} likely underestimates the acceleration of the infalling matter, and thus the timescale of the explosion. With a longer timescale, sub-dominant reactions, such as $3\alpha$ and $^{12}$C($\alpha,\gamma$)$^{16}$O can proceed and the resultant $^{16}$O can then fuel further explosive alpha process reactions. From this viewpoint, it is natural to expect the total energy produced to be several times greater than in our calculations where the $3\alpha$ and $^{12}$C($\alpha,\gamma$)$^{16}$O do not meaningfully contribute. We also note that in their simulation, the carbon mass does not change ($\Delta ^{12}$C = -3 $\msun$, Table 1 of \citealt{chen2014}) which may indicate that $3\alpha$ and $^{12}$C($\alpha,\gamma$)$^{16}$O do not in fact occur, contrary to what is written in the text. If this is the case, then the explanation for the discrepancy in total energy may be simply that the shorter timescale of our explosion leads to less nuclear burning (though using the same reactions), along with other factors such as the difference in max $T_c$, our fully relativistic code, and our smaller progenitor mass. 

The GRSNe in this work also have lower energies than the one found in \citet{nagele2020}. This is due to the lower progenitor mass and to the catalysis discussed in Sec. \ref{methods_HYDnuc}. Despite the differences in explosion energy, the composition of our explosion ejecta is similar to previous works (Table \ref{tab:ejecta}). The main difference between the GRSN in \citet{nagele2020} and the GRSNe in this work is the wider mass range and the discovery of pulsations, which widens the mass range for observations even further.

\citet{moriya2021} found that the observational duration of a GRSN will be on the order of $10^{2-3}$ seconds for the peak emission and $10^{2-3}$ days for the plateau, which shares some similarities with Type IIP SNe. Because the GRSN would have to occur in the high redshift universe, this means that the observer duration is longer by about a factor of ten. \citet{moriya2021} demonstrated that the GRSN plateau may be differentiated from other persistent sources if it is observed in multiple bands. \citet{whalen2013} studied the same GRSN using different spectral codes. They also assume two different circum-stellar media, one being a wind driven by the SMS and the other being in-falling matter. In the former case, they find that the emission could last 400 to 1000 days. They also find that stars at redshift 30 will be clearly visible to JWST. In the latter case, the emission is longer and more sporadic, with the observer duration stretching from roughly 1000 to 4000 days. In the future, we plan to investigate if the lower energy of our GRSN would significantly alter any of these findings.

We determined that none of the observed metal poor stars match the inferred abundance pattern from the GRSNe, and from this we conclude that none of these stars were singly enriched by a GRSN. However, we note that the multi-enrichment scenario cannot be ruled out.

As in all numerical studies, there are numerous uncertainties. In HOSHI, along with the usual sources of numerical error, we also need to consider the error due to using the post Newtonian approximation. After the inclusion of internal energy to density (Eq. \ref{eq:epsilon}) in the TOV equation, the post Newtonian pressure gradient matches the TOV pressure gradient to one part in $10^5$. Although this level of accuracy likely supersedes numerical error, we have to keep in mind that the TOV equation assumes a hydrostatic configuration, which, for instance, is not true as the star contracts towards the end of hydrogen burning. Finally, and perhaps most importantly, our evolutionary models are not rotating, whereas real SMSs are expected to initially be medium rotators \citep{hammerle2018} which may spin up over the course of their lifetimes \citep{maeder2001}. We intend to more fully investigate the effects of rotation in the future. Finally, we would like to point out that this paper has focused exclusively on the GR radial instability, but it is also possible that the SMS could experience other instabilities earlier in its lifetime. 

Regarding the GR stability analysis, we were able to quantify the error of $|\omega_0^2|$ for polytropes (Fig. \ref{fig:nconv_inst}) as being around $10^{-7,-8}$. As seen in Fig. \ref{fig:insta_hoshi}, typical values of $\omega_0^2$ are greater than this error. Although we demonstrated that the error decreases for increasing resolution (Fig. \ref{fig:nconv_inst}) the gain in accuracy is low compared to the gain in computational time that an increased mesh point number of an order of magnitude or two would require. 

As far as sources of error in HYDnuc, \citet{chen2014} showed that multidimensional effects may not play a big role in the GRSN (besides increasing the explosion energy), and they are also not thought to contribute to the instability \citep{chandrasekhar1965}. We verified that radiative and convective energy transport do not effect the explosion outcome and tested that the explosion energy does not depend on the choice of the carbon alpha capture rate.

In the future, we plan to apply the GR stability analysis to rotating SMSs, as well as verifying our current results with multidimensional simulations. We also intend to asses the observability of the GRSN using methods similar to those in \citet{moriya2021}. Finally, we will investigate the possibility of multiple pulsations.

\section*{Data Availability}

The data underlying this article will be shared on reasonable request to the corresponding author.

\section*{Acknowledgements}

We thank the anonymous referee for their extensive and careful comments which greatly improved the final manuscript. This study was supported in part by the Grant-in-Aid for the Scientific Research of Japan Society for the Promotion of Science (JSPS, Nos. JP19K03837, JP20H01905, JP20H00158, JP21H01123) and by Grant-in-Aid for Scientific Research on Innovative areas (JP17H06357, JP17H06365) from the Ministry of Education, Culture, Sports, Science and Technology (MEXT), Japan. For providing high performance computing resources, YITP, Kyoto University is acknowledged. K.S. would also like to acknowledge computing resources at KEK and RCNP Osaka University. C.N. would like to thank Tilman Hartwig for discussions of metal poor stars and Masao Takata for discussions of stellar oscillations.




\bibliographystyle{mnras}
\bibliography{bib}




\section{Appendix A: Iterative method for solving the perturbation equations}

As mentioned in the text, we adopt a straightforward numerical approach to solving the perturbation equation (Eqs. \ref{eq_N}, \ref{eq_GR}). From the stellar evolution calculation, we take a stellar profile (for a particular time step), so that the only unknowns in Eqs. \ref{eq_N}, \ref{eq_GR} are $\xi$ and $\omega$. Our aim, then, is to find a perturbation $\xi$ and its associated frequency $\omega$ which satisfy the equations for the particular stellar structure. Note that the below discussion can be applied either to Eq. \ref{eq_N} or Eq. \ref{eq_GR}.

First an initial guess is made for $\omega_0^2$, and the equation is spatially integrated once from the inner boundary $\xi_{\rm inner}$ ($\xi(0)=0$, $\xi'(0)=1$) and once from the outer boundary $\xi_{\rm outer}$ ($\xi(R)=1$, $\xi'(R)=0$). The integration is performed using second order Euler's method, where the simplicity of the method is due to the limited number of grid points in the stellar evolution calculation. Finally, the values of $\xi'(0)$ and $\xi(R)$ should be non zero. 

Next, we compare $\xi_{\rm inner}$ to $\xi_{\rm outer}$ at some radius $p$ where $0<p<R$ using the Wronskian of the two solutions normalized by their amplitude:

\begin{equation}
     \mathcal{X}(p) = \frac{2\mathcal{W}(p)}{\xi_{\rm inner}(p)+\xi_{\rm outer}(p)} 
\end{equation}
where we have divided by the amplitude in order to prevent $\xi$ from becoming too small. In general, $p$ may be chosen freely though for numerical models with finite resolution it should not be too close to either boundary. We choose $p$ to be at min$|\mathcal{X}(r)|$.

The goal is to find $\xi_{\rm inner}$, $\xi_{\rm outer}, \omega_0^2$ such that $|\mathcal{X}(p) < \mathcal{T}|$, where $\mathcal{T}$ is a threshold we set at $10^{-10}\times \xi'(0)$ for $\omega_0^2>0$ and $10^{-5}\times \xi'(0)$ for $\omega_0^2<0$ (the latter being smaller due to computational costs). This is accomplished iteratively by linear extrapolation of $\mathcal{X}(p)$ in the space of $\omega_0^2$ for two sets of $\xi_{\rm inner}$, $\xi_{\rm outer}, \omega_0^2$.

If a solution is not found within a set number of iterations, or if a solution is found for a higher order mode (for instance if it finds $\omega^2_1$ instead of $\omega^2_0$), we redo the calculation with a different value of $\xi'(0)$. This step circumvents numerical overflow which may occur if our initial guess of the slope is wrong. This step only becomes necessary when $\xi$ is extremely nonlinear, which may occur for $\omega_0^2 \ll 0$.

\begin{figure*}
    \centering
    \includegraphics[width=2\columnwidth]{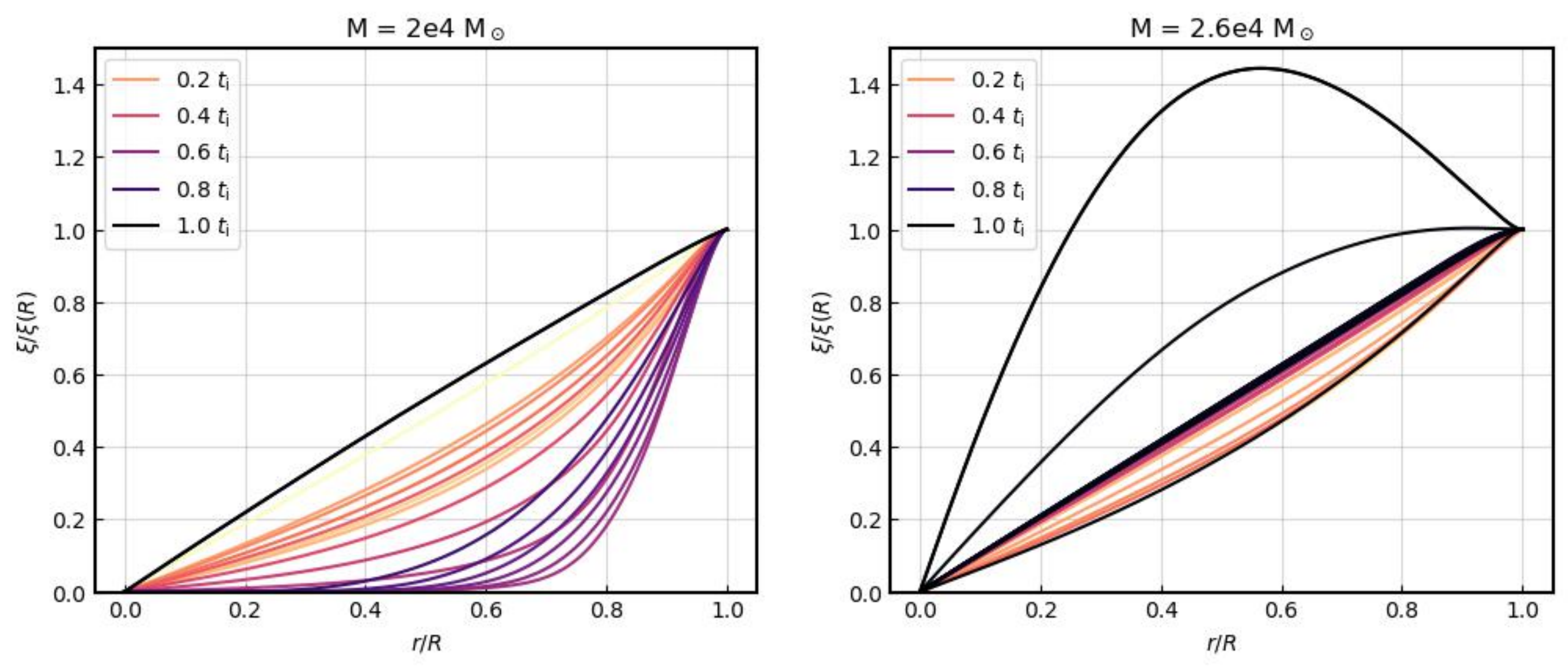}

    \caption{Normalized amplitude of the fundamental mode of the perturbation at various time snapshots leading up to the first instability (denoted by $t_i$). The left panel shows the 2 $\times 10^4$ $\msun$ model which has a nearly linear perturbation at $t_i$ while the right panel shows 2.6 $\times 10^4$, which has more amplitude concentrated at smaller radius at $t_i$. The perturbations have been normalized to $\xi_0 (R) = 1$.}
    \label{fig:xi_r}
\end{figure*}

\begin{figure*}
    \centering
    \includegraphics[width=2\columnwidth]{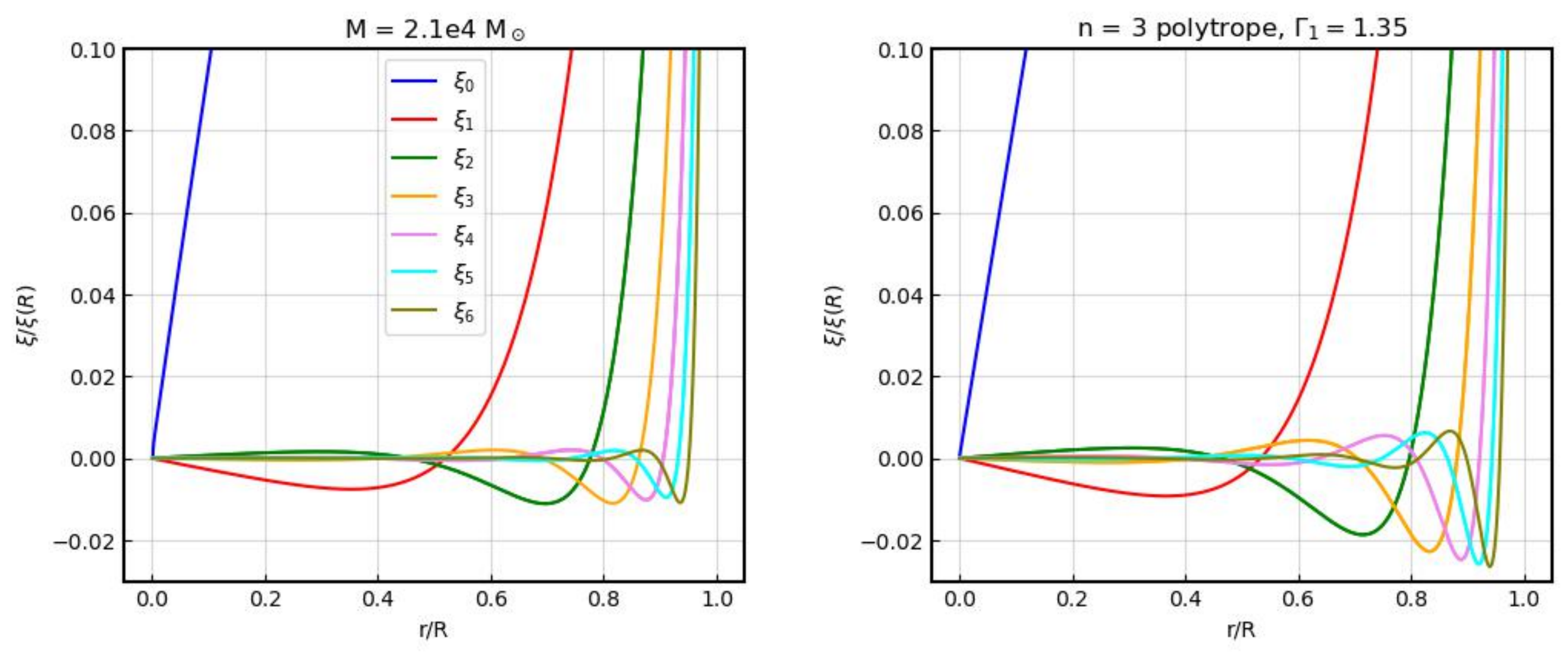}

    \caption{First six modes of the perturbation for the first timestep in the 2.1 $\times 10^4$ $\msun$ calculation (left panel) and an $n=3$ polytrope (right panel). The perturbations have been normalized to $\xi_n (R) = 1$.}
    \label{fig:xi_n}
\end{figure*}

Figs. \ref{fig:xi_r}, \ref{fig:xi_n} show examples of the perturbation found using this method. In Fig. \ref{fig:xi_r}, we show the time evolution of the fundamental mode of the perturbation ($\xi_0$) from the beginning of the HOSHI simulation until the first instability. In Fig. \ref{fig:xi_n}, we show the first six modes of the perturbation for a SMS and for a polytrope, which are noticeably different from one another. Our choices for $\mathcal{T},\mathcal{X}$ are somewhat arbitrary, but we have at least demonstrated their efficacy with tests on numerical polytropes (Sec. \ref{methods_stab}).

\section{Appendix B: Steady state calculation}

In this section, we will compare the 58 and 61 isotope networks (Table \ref{tab:networks}), the latter of which contains aluminum isotopes and isomers allowing magnesium to absorb excess protons, which in turn prevents an unrealistic enhancement of the carbon alpha capture rate. 

\begin{figure}
    \centering
    \includegraphics[width=\columnwidth]{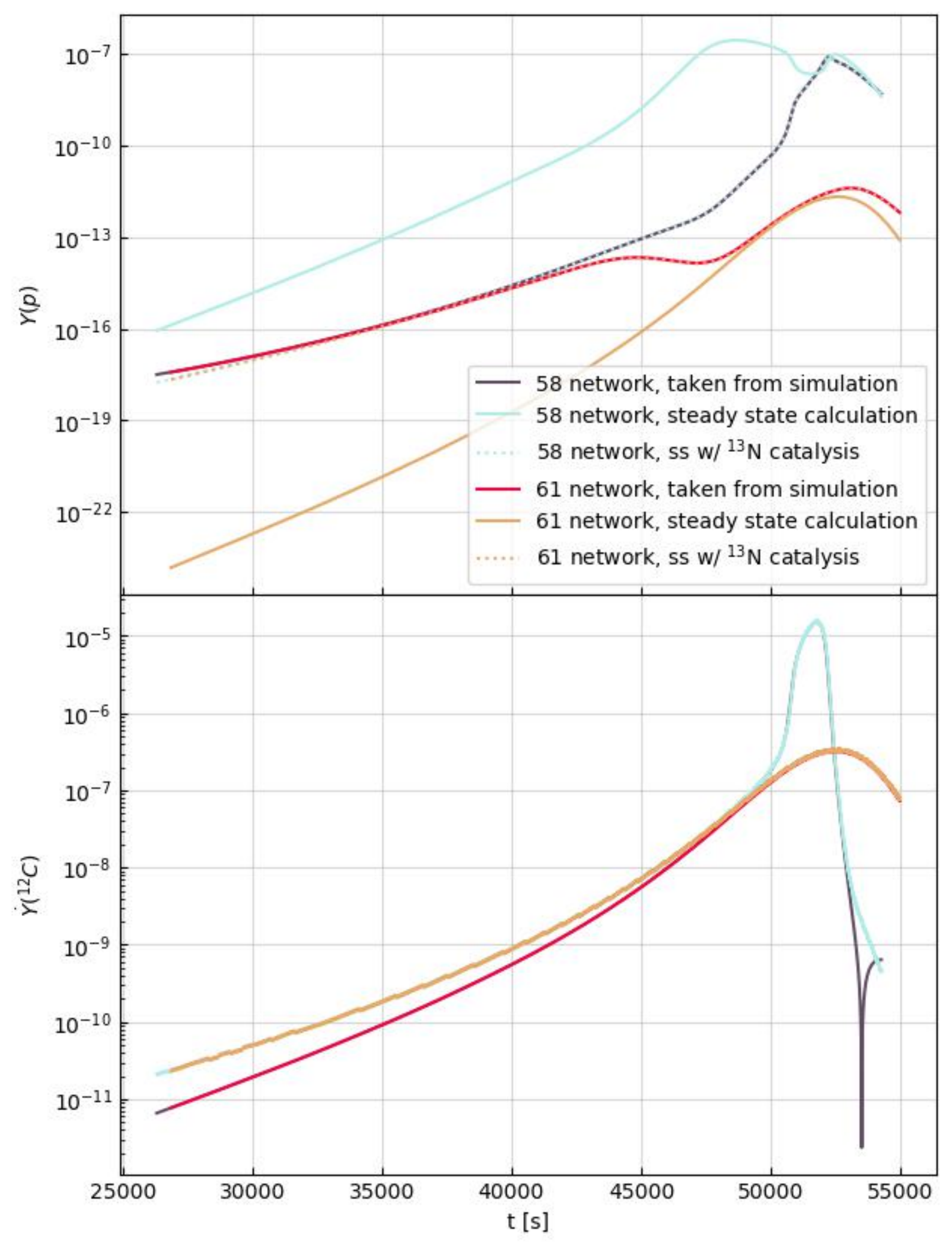}

    \caption{Upper panel --- proton number fraction from the steady state calculation, and from the simulation, for the 58 and 61 isotope networks. Dotted lines include the $^{13}$N catalysis reactions. Lower panel --- time derivative of carbon number fraction. }
    \label{fig:ss}
\end{figure}

Assume a steady state for the number fraction of protons
\begin{equation}
    \nonumber
    \dot{Y}(p) = 0.
\end{equation}
Ignoring the $^{13}$N catalysis reactions for now, the proton number changes as 
\begin{equation}
    \nonumber
    \dot{Y}(p) = \Lambda_{^{24}Mg(\alpha,p)^{27}Al} Y(^{24}Mg)Y(\alpha) - \Lambda_{^{27}Al(p,\gamma)^{28}Si} Y(^{27}Al)Y(p) 
\end{equation}
\begin{equation}    
    \nonumber
    -\Lambda_{^{24}Mg(p,\gamma)^{25}Al} Y(^{24}Mg)Y_{61}(p) + ...
\end{equation}
where we have written $Y_{61}(p)$ in the third reaction to show that this reaction only occurs in the 61 isotope network. We can solve for $Y(p)$:
\begin{equation}
    \nonumber
    Y(p)  = \frac{\Lambda_{^{24}Mg(\alpha,p)^{27}Al} Y(^{24}Mg)Y(\alpha) + ...}{\Lambda_{^{27}Al(p,\gamma)^{28}Si} Y(^{27}Al) + \Lambda_{^{24}Mg(p,\gamma)^{25}Al} Y(^{24}Mg)Y_{61}(p) + ... }
\end{equation}
so Y(p) will be smaller in the 61 network calculation by a factor of 
\begin{equation}
    \nonumber
    \frac{Y_{61}(p)}{Y_{58}(p)}= 
    \frac{\Lambda_{^{27}Al(p,\gamma)^{28}Si} Y(^{27}Al) + ... }{\Lambda_{^{27}Al(p,\gamma)^{28}Si} Y(^{27}Al) + \Lambda_{^{24}Mg(p,\gamma)^{25}Al} Y(^{24}Mg)Y_{61}(p) + ...}
\end{equation}
which is order $10^{-5}$ (Fig. \ref{fig:ss}). Including the above reactions gives us the correct order of magnitude, but additionally including the $^{13}$N catalysis reactions would give the precise behaviour, as can be seen by the dotted lines in Fig. \ref{fig:ss}. Next, consider carbon, 
\begin{equation}
    \nonumber
    \dot{Y}(^{12}C)  = \Lambda_{^{12}C(\alpha,\gamma)^{16}O} Y(^{12}C)Y(\alpha) + \Lambda_{^{12}C(p,\gamma)^{13}N} Y(^{12}C) Y(p) + ...
\end{equation}
where the only difference between the two networks is $Y(p)$. 
\begin{equation}
    \nonumber
    \frac{\dot{Y}_{61}(^{12}C)}{\dot{Y}_{58}(^{12}C)}  = \frac{\Lambda_{^{12}C(\alpha,\gamma)^{16}O} Y(^{12}C)Y(\alpha) + \Lambda_{^{12}C(p,\gamma)^{13}N} Y(^{12}C) Y_{61}(p) + ...}{\Lambda_{^{12}C(\alpha,\gamma)^{16}O} Y(^{12}C)Y(\alpha) + \Lambda_{^{12}C(p,\gamma)^{13}N} Y(^{12}C) Y_{58}(p) + ..} 
\end{equation}
which is order $10^{-2}$ (Fig. \ref{fig:ss}). So, the 58 isotope network overestimates carbon burning by a factor of 100, which significantly alters the course of the simulation.

\begin{table}
	\centering
	\caption{Summary table for nuclear networks. Entries show the range in A for the specified element.}
	\label{tab:networks}
	\begin{tabular}{|c|l|l|l|l|l|l|l|} 
		
    		Element & 52 & 58 & 61 & 79 & 89 & 153 & 300 \\
    		\hline

n&1&1&1&1&1&1&1\\
p&1-3&1-3&1-3&1-3&1-3&1-3&1-3\\
He&3-4&3-4&3-4&3-4&3-4&3-4&3-4\\
Li&6-7&6-7&6-7&6-7&6-7&6-7&6-7\\
Be&7-9&7-9&7-9&7-9&7-9&7-9&7-9\\
B&8-11&8-11&8-11&8-11&8-11&8-11&8-11\\
C&12-13&12-13&12-13&12-13&12-13&12-13&11-16\\
N&13-15&13-15&13-15&13-15&13-15&13-15&13-18\\
O&14-18&14-18&14-18&14-18&14-18&14-18&14-20\\
F&17-19&17-19&17-19&17-19&17-19&17-19&17-22\\
Ne&18-20&18-20&18-20&18-22&18-22&18-22&18-24\\
Na&23&23&23&23&21-23&21-23&21-26\\
Mg&24&24-26&24-26&22-26&22-26&22-26&22-28\\
Al&27&27&25-27&25-27&25-27&25-27&25-30\\
Si&28&28-30&28-30&26-32&26-32&26-32&26-32\\
P&31&31&31&31&29-33&29-33&27-34\\
S&32&32-34&32-34&30-36&30-36&30-36&30-37\\
Cl&35&35&35&35&33-37&33-37&32-38\\
Ar&36&36&36&34-40&34-40&34-40&34-43\\
K&39&39&39&39&39&37-41&36-45\\
Ca&40&40&40&40&40&38-43&38-48\\
Sc&43&43&43&43&43&41-45&40-49\\
Ti&44&44&44&44&44&43-48&42-51\\
V&47&47&47&47&47&45-51&44-53\\
Cr&48&48&48&48&48&47-54&46-55\\
Mn&51&51&51&51&51&49-55&48-57\\
Fe&52-56&52-56&52-56&52-56&52-56&51-58&50-61\\
Co&55-56&55-56&55-56&55-56&55-56&53-59&51-62\\
Ni&56&56&56&56&56&55-62&54-66\\
Cu&---&---&---&---&---&57-63&56-68\\
Zn&---&---&---&---&---&60-64&59-71\\
Ga&---&---&---&---&---&---&61-73\\
Ge&---&---&---&---&---&---&63-75\\
As&---&---&---&---&---&---&65-76\\
Se&---&---&---&---&---&---&67-78\\
Br&---&---&---&---&---&---&69-79\\

		\hline
	\end{tabular}
\end{table}

\begin{table}
	\centering
	\caption{Yields of all isotopes in the network in units of $\msun$ for each of the two GRSNe. Full table available online.}
	\label{tab:networks}
	\begin{tabular}{|c|l|l|} 
		
    		Isotope & 2.95 $\times$ $10^4$ $\msun$ & 3.00 $\times$ $10^4$ $\msun$ \\
    		\hline

        n    &     3.269403297e-45  &   7.494798495e-34 \\
        p    &     5.441498563e+03  &   5.536626361e+03 \\
        d    &     3.420657731e-14  &   8.386696275e-11 \\
        t    &     1.536980840e-22  &   3.756248063e-15 \\
      he3    &     1.136117283e-02  &   1.218897594e-02 \\
      he4    &     1.694646913e+04  &   1.807698033e+04 \\
      li6    &     2.101344385e-15  &   4.646143231e-13 \\
      li7    &     2.921447296e-07  &   3.798748682e-07 \\
      ...    &     ...   &   ...  \\

		\hline
	\end{tabular}
\end{table}

\bsp	
\label{lastpage}
\end{document}